\newcommand\beq{\begin{eqnarray}}
\newcommand\eeq{\end{eqnarray}}

\newcommand\ETmiss{E_T^{\rm miss}}

\newcommand\Nzero{N}
\newcommand\Nplus{V}
\newcommand\Nminus{U}
\newcommand\nzero{\widehat N}
\newcommand\nplus{\widehat V}
\newcommand\nminus{\widehat U}
\newcommand\DeltaMappendix{Appendix A}
\newcommand\STappendix{Appendix B}
\newcommand\Decayappendix{Appendix C}
\newcommand\Crosssectionappendix{Appendix D}
\newcommand\epsT{\epsilon_T}
\newcommand\epsL{\epsilon_L}
\newcommand\yk{k}
\newcommand\yh{k'}

\def\lsim{\mathrel{\rlap{\lower4pt\hbox{$\sim$}}
    \raise1pt\hbox{$<$}}}                
\def\gsim{\mathrel{\rlap{\lower4pt\hbox{$\sim$}}
    \raise1pt\hbox{$>$}}}

\documentclass[
amsmath,
prd,nofootinbib,floatfix,11pt,
]{revtex4}

\allowdisplaybreaks
\usepackage{graphicx}
\usepackage{setspace}

\begin{document}
\renewcommand{\theequation}{\arabic{section}.\arabic{equation}}

\title{\Large%
\baselineskip=21pt
Raising the Higgs mass with Yukawa couplings for isotriplets\\
in vector-like extensions of minimal supersymmetry}

\author{Stephen P. Martin}
\affiliation{
{\it Department of Physics, Northern Illinois University, DeKalb IL 60115,} 
and 
\\
{\it Fermi National Accelerator Laboratory, P.O. Box 500, Batavia IL 60510.}}

\begin{abstract}\normalsize 
\baselineskip=15pt 
Extra vector-like matter with both electroweak-singlet masses and 
large Yukawa couplings can significantly raise the lightest Higgs boson
mass in supersymmetry through radiative corrections. I consider models of
this type that involve a large Yukawa coupling between weak isotriplet and
isodoublet chiral supermultiplets. The particle content can be completed
to provide perturbative gauge coupling unification, in several different
ways. The impact on precision electroweak observables is shown to be 
acceptably small, even if the new particles are as light as the current
experimental bounds of order 100 GeV. I study the corrections to the 
lightest Higgs boson mass, and discuss the general features of the collider
signatures for the new fermions in these models.
\end{abstract}


\maketitle

\baselineskip=14.9pt
\tableofcontents

\vfill\eject
\baselineskip=14.5pt

\setcounter{footnote}{1}
\setcounter{page}{2}
\setcounter{figure}{0}
\setcounter{table}{0}

\section{Introduction}
\label{sec:intro}
\setcounter{equation}{0}
\setcounter{footnote}{1}

Supersymmetry as an extension of the Standard Model addresses the 
hierarchy problem associated with the small ratio of the electroweak 
breaking scale to the Planck scale or other very high energy scales. 
However, the lack of a signal for the lightest neutral scalar boson, 
$h^0$, at the CERN LEP2 $e^+e^-$ collider imposes some tension on the 
minimal supersymmetric standard model (MSSM) parameter space, motivating 
an examination of further extensions that can increase the theoretical 
prediction of the mass of $h^0$.

In minimal supersymmetry, the biggest radiative corrections to $m^2_{h^0}$ 
come from one-loop diagrams with top quarks and squarks, and are 
proportional to the fourth power of the top Yukawa coupling. This 
suggests that one could improve the situation by introducing new 
supermultiplets with large Yukawa couplings that would raise the $h^0$ 
mass still further. This has been considered for the case of a fourth 
chiral family \cite{Fok:2008yg,Litsey:2009rp}. However, in supersymmetry, 
the Yukawa couplings of a fourth chiral family would have to be so large 
(in order to evade discovery by LEP2 and the Tevatron) that perturbation 
theory would break down not far above the electroweak scale 
\cite{Fok:2008yg}. This would mean that the apparent success of gauge 
coupling unification in the MSSM is merely an illusion. Even accepting this, 
the corrections to precision electroweak physics would be too large in 
most of the parameter space, unless there are rather specific splittings of 
fermion masses \cite{Kribs:2007nz}.

Instead, one can consider models with extra matter in chiral 
supermultiplets comprised of vector-like representations of the Standard 
Model gauge group $SU(3)_c \times SU(2)_L \times U(1)_Y$, i.e., those 
that allow tree-level superpotential mass terms before spontaneous 
electroweak symmetry breaking. These bare mass terms are responsible for 
most of the vector-like fermion masses. However, if the extra vector-like 
matter includes appropriate representations differing by $1/2$ unit of 
weak isospin, then they can also have Yukawa couplings to the MSSM Higgs 
supermultiplets. If large enough, these new Yukawa couplings can yield a 
significant enhancement of 
$m_{h^0}^2$ through one-loop effects, helping to explain why $h^0$ was not 
kinematically accessible to LEP2.

Earlier model-building work 
\cite{Moroi:1991mg,Babu:2004xg,Babu:2008ge,Martin:2009bg,Graham:2009gy} 
along these lines has considered 
extra vector-like matter transforming in gauge representations of 
the types already present in the MSSM, and their conjugates. Under 
$SU(3)_C \times SU(2)_L \times U(1)_Y$, these candidate extra superfields 
transform like:
\beq
&&
Q = ({\bf 3}, {\bf 2}, 1/6), 
\qquad\>
\overline{Q} = ({\bf \overline{3}}, {\bf 2}, -1/6), 
\qquad
U = ({\bf 3}, {\bf 1}, 2/3), 
\qquad
\overline{U} = ({\bf \overline{3}}, {\bf 1}, -2/3), 
\nonumber
\\
&&D = ({\bf 3}, {\bf 1}, -1/3), 
\qquad\>\!\!\!\!\!
\overline{D} = ({\bf \overline{3}}, {\bf 1}, 1/3), 
\qquad\>\>\,
L = ({\bf 1}, {\bf 2}, -1/2), 
\qquad\!\!\!
\overline{L} = ({\bf 1}, {\bf 2}, 1/2), 
\nonumber
\\
&&
E = ({\bf 1}, {\bf 1}, -1), 
\qquad\,\>
\overline{E} = ({\bf 1}, {\bf 1}, 1), 
\qquad\>\>\>\>\>\>\>
N, \overline{N} = ({\bf 1}, {\bf 1}, 0).
\label{eq:extraslist}
\eeq
(Each bar appearing here is part of the name of the field, and does not 
denote any kind of conjugation.) Requiring that these new particles are 
not much heavier than 1 TeV, and that the gauge couplings still unify 
perturbatively, there are three types of models, with new (non-MSSM) 
chiral supermultiplets:
\beq
\mbox{LND$^n$ models:}\qquad\>&&
n\times (L, \overline L, N,  \overline N, D, \overline D)\qquad \quad 
{\rm for}\>\,(n=1,2,3), 
\label{eq:defLND}
\\
\mbox{QUE model:}\,\qquad\>\>
&&
Q, \overline Q, U, \overline U, E, \overline E,
\label{eq:defQUE}
\\
\mbox{QDEE model:}\qquad\>\>
&&
Q, \overline Q, D, \overline D, E, \overline E, 
E, \overline E.
\label{eq:defQDEE}
\eeq
In each case, the number of singlets $N$ or $\overline N$ is actually 
arbitrary, since they do not directly affect the running of the gauge 
couplings, but including the $N,\overline N$ in the (LND)$^n$ models 
allows new Yukawa couplings. There is also a possible model with new 
supermultiplets:
\beq
\mbox{QUDLE model:}\qquad
Q, \overline Q, U, \overline U, D, \overline D, L, 
\overline L, E, \overline E.
\eeq
However, to avoid the gauge couplings become non-perturbative in the 
ultraviolet before they have a chance to unify,\footnote{To correctly 
implement this perturbativity requirement, it is mandatory to use 2-loop 
(or higher) beta functions. The numerical results in this paper always 
use 2-loop beta functions for all parameters. These can be obtained 
straightforwardly from the general results listed in 
\cite{betas:1,betas:2}, and so are not listed explicitly here.} the 
average of the new particle masses in the QUDLE model would have to be at 
least about 2.5 TeV. This does not rule out the QUDLE model, but it goes 
strongly against the motivation of avoiding fine tuning. (If the large 
masses of the new fermions are due mostly to supersymmetric mass terms, 
then one cannot have a large enough hierarchy between scalar and fermion 
masses to increase $m^2_{h^0}$ appreciably, unless the soft supersymmetry 
breaking scalar masses are much larger still.) Up to the inclusion of 
singlets, the LND model content corresponds to a ${\bf 5} + {\bf 
\overline{5}}$ of $SU(5)$, the QUE model to a ${\bf 10} + {\bf 
\overline{10}}$ of $SU(5)$, and the QUDLE model to a ${\bf 16} + {\bf 
\overline{16}}$ of $SO(10)$, although one need not subscribe to a belief 
in those groups as grand unified gauge symmetries.

In ref.~\cite{Martin:2009bg}, I showed that the LND, QUE and QDEE models 
are compatible with precision electroweak constraints, even if the new 
Yukawa couplings are as large as their quasi-fixed-point values and the 
new quarks and leptons are approximately as light as their present direct 
search limits from Tevatron and LEP2.

However, the new vector-like matter may include other representations not 
listed in eq.~(\ref{eq:extraslist}). Let us denote possible $SU(2)_L$ 
triplet and $SU(3)_C$ octet chiral supermultiplets by:
\beq
T = ({\bf 1}, {\bf 3}, 0), 
\qquad
O = ({\bf 8}, {\bf 1}, 0).
\label{eq:TO}
\eeq
These are real representations of the gauge group, and so can have 
Majorana-type superpotential mass terms by themselves. If we denote by 
$n_Q$ the number of $Q,\overline Q$ pairs, and similarly for $n_U$, 
$n_D$, $n_L$, and $n_E$, and denote by $n_T$ and $n_O$ the number of $T$ 
and $O$ supermultiplets respectively, then the one-loop beta functions 
for the gauge couplings (with a GUT normalization $g_1 = \sqrt{5/3} g'$)
are:
\beq
Q\frac{dg_1}{dQ} = \beta_{g_1} &=& \frac{g_1^3}{16\pi^2} 
\bigl (33 + n_Q + 8 n_U + 2 n_D + 3 n_L + 6 n_E \bigr )/5
,
\\
Q\frac{dg_2}{dQ} =
\beta_{g_2} &=& \frac{g_2^3}{16\pi^2} \bigl ( 1 + 3 n_Q + n_L + 2 n_T \bigr )
,
\\
Q\frac{dg_3}{dQ} =\beta_{g_3} &=& \frac{g_3^3}{16\pi^2} 
\bigl ( -3 + 2 n_Q + n_U + n_D + 3 n_O \bigr )
,\eeq
where $Q$ is the renormalization scale.
Perturbative unification requires that the one-loop contributions to the 
beta functions from the new fields are equal and not too large, so that 
\beq
(n_Q + 8 n_U + 2 n_D + 3 n_L + 6 n_E \bigr )/5
= 3 n_Q + n_L + 2 n_T  = 2 n_Q + n_U + n_D + 3 n_O \equiv N,
\label{eq:defN}
\eeq
where $N$ is 1, 2, or 3. (The details and precise quality of the 
unification depend also on 2-loop effects, including the effects 
of new Yukawa couplings. However, these effects do not make a dramatic 
difference, provided that $N\leq 3$.) This allows us to recognize some 
model possibilities different from those in 
eqs.~(\ref{eq:defLND})-(\ref{eq:defQDEE}). Consider models with 
extra chiral supermultiplets beyond the MSSM:
\beq
\mbox{TUD model:}\qquad
&&T,U,\overline{U},D,\overline{D}, \\
\mbox{TEDD model:}\qquad
&&T,E,\overline{E},
D,\overline{D},D,\overline{D},\\
\mbox{OLLLE model:}\qquad
&&O,L,\overline{L},L,\overline{L},L,\overline{L},E,\overline{E},
\\
\mbox{OTLEE model:}\qquad
&&O,T,L,\overline{L},E,\overline{E},E,\overline{E},
\label{OTLEE}\\
\mbox{TLUDD model:}\qquad
&&T,L,\overline{L},U,\overline{U},D,\overline{D},D,\overline{D},\\
\mbox{TLEDDD model:}\qquad
&&T,L,\overline{L},E,\overline{E},D,\overline{D},D,\overline{D},D,\overline{D}.
\eeq
The first two have $N=2$, and the last four have $N=3$.
As before, these models can be augmented by any number of gauge singlet 
supermultiplets,\footnote{For example, the OTLEE model 
augmented by five singlets would correspond to an adjoint 
representation of the GUT group $SU(3)_c \times SU(3)_L \times 
SU(3)_R \subset E_6$.} which do not affect the gauge coupling running.

There are other possible representations
of $SU(3)_c \times SU(2)_L \times U(1)_Y$ that one could try to include.
However, if one requires no unconfined fractional electric charges, then 
all such vector-like combinations, which include for example
$({\bf 1}, {\bf 5}, 0)$ or
$({\bf 1}, {\bf 3}, 1) + ({\bf 1}, {\bf 3}, -1)$ or
$({\bf 1}, {\bf 2}, 3/2) + ({\bf 1}, {\bf 2}, -3/2)$ or
$({\bf 3}, {\bf 2}, -5/6) + ({\bf \overline{3}}, {\bf 2}, 5/6)$ or
$({\bf 3}, {\bf 1}, -4/3) + ({\bf \overline{3}}, {\bf 1}, 4/3)$ or
$({\bf 6}, {\bf 1}, 1/3) + ({\bf \overline{6}}, {\bf 1}, -1/3)$ or
$({\bf 1}, {\bf 1}, 2) + ({\bf 1}, {\bf 1}, -2)$,
would contribute too much to $N$ and can not be consistent with 
perturbative gauge coupling unification, unless the average of the new 
particle masses is at least in the multi-TeV range. So with these 
requirements, $T$ and $O$ are the only new possibilities beyond 
eq.~(\ref{eq:extraslist}). Restricting the new supermultiplets to those 
in eqs.~(\ref{eq:extraslist}) and (\ref{eq:TO}) assures that small 
mixings with the MSSM quark and lepton or gaugino and higgsino fields can 
eliminate stable exotic particles, which otherwise could be disastrous 
relics of the early universe.
For some other recent discussions of vector-like supermultiplets in 
supersymmetry, see \cite{Liu:2009cc}-\cite{Li:2010hi}.

In this paper, I will reserve the capital letters as above for new extra 
supermultiplets, and use lowercase letters for the usual chiral MSSM 
quark and lepton supermultiplets:
\beq
&&
q_i = ({\bf 3}, {\bf 2}, 1/6), 
\qquad
\overline{u}_i = ({\bf \overline{3}}, {\bf 1}, -2/3), 
\qquad
\overline{d}_i = ({\bf \overline{3}}, {\bf 1}, 1/3), 
\qquad
\ell_i = ({\bf 1}, {\bf 2}, -1/2), 
\nonumber \\ &&
\overline{e}_i = ({\bf 1}, {\bf 1}, 1), 
\qquad
H_u = ({\bf 1}, {\bf 2}, 1/2), 
\qquad
H_d = ({\bf 1}, {\bf 2}, -1/2),
\eeq
with $i = 1,2,3$ denoting the three families.
The MSSM part of the superpotential, in the approximation that only 
third-family Yukawa couplings are included, is:
\beq
W = \mu H_u H_d + y_t \overline u_3 q_3  H_u - y_b \overline d_3 q_3 H_d 
- y_\tau \overline e_3 \ell_3 H_d.
\label{eq:WMSSM}
\eeq
[Products of weak isospin doublet fields implicitly have their $SU(2)_L$ 
indices contracted with an antisymmetric tensor $\epsilon^{12} = 
-\epsilon^{21}$, with the first component of every doublet having $T_3 = 
1/2$ and the second component having $T_3 = -1/2$. So, for example $q_3 
H_d= t H_d^- - b H_d^0$, with the minus signs working out to give 
positive masses after the neutral components of the Higgs fields get 
vacuum expectation values (VEVs).]

Because of their vector-like representations, any Yukawa coupling-induced 
mixing between the new fields $Q,\overline Q, U, \overline U, D, 
\overline D, L, \overline L, E, \overline E$ and their MSSM counterparts 
will not be governed by a GIM mechanism, and so must be highly 
suppressed. Therefore, to first approximation one can consider only 
Yukawa couplings that connect pairs of new fields. This can be enforced 
by an (approximate) symmetry, for example a $Z_2$ under which the new 
superfields are odd and the MSSM quark and lepton superfields are even, 
or vice versa. The TUD and TEDD models do not have any allowed 
Yukawa couplings between pairs of new fields, and the OLLLE 
model allows only Yukawa couplings of the form $H_u \overline L E$ and 
$H_d L \overline E$ (and $H_u L \overline N$ and $H_d \overline L 
N$ if singlets are present), which are qualitatively similar to the ones in the 
LND model already studied in refs.~\cite{Babu:2008ge,Martin:2009bg}, with 
fixed points that are not large enough to raise the $h^0$ mass by a very 
significant amount.

\begin{figure}[!tp]
\begin{minipage}[]{0.52\linewidth}
\begin{flushleft}
\includegraphics[width=7.0cm,angle=0]{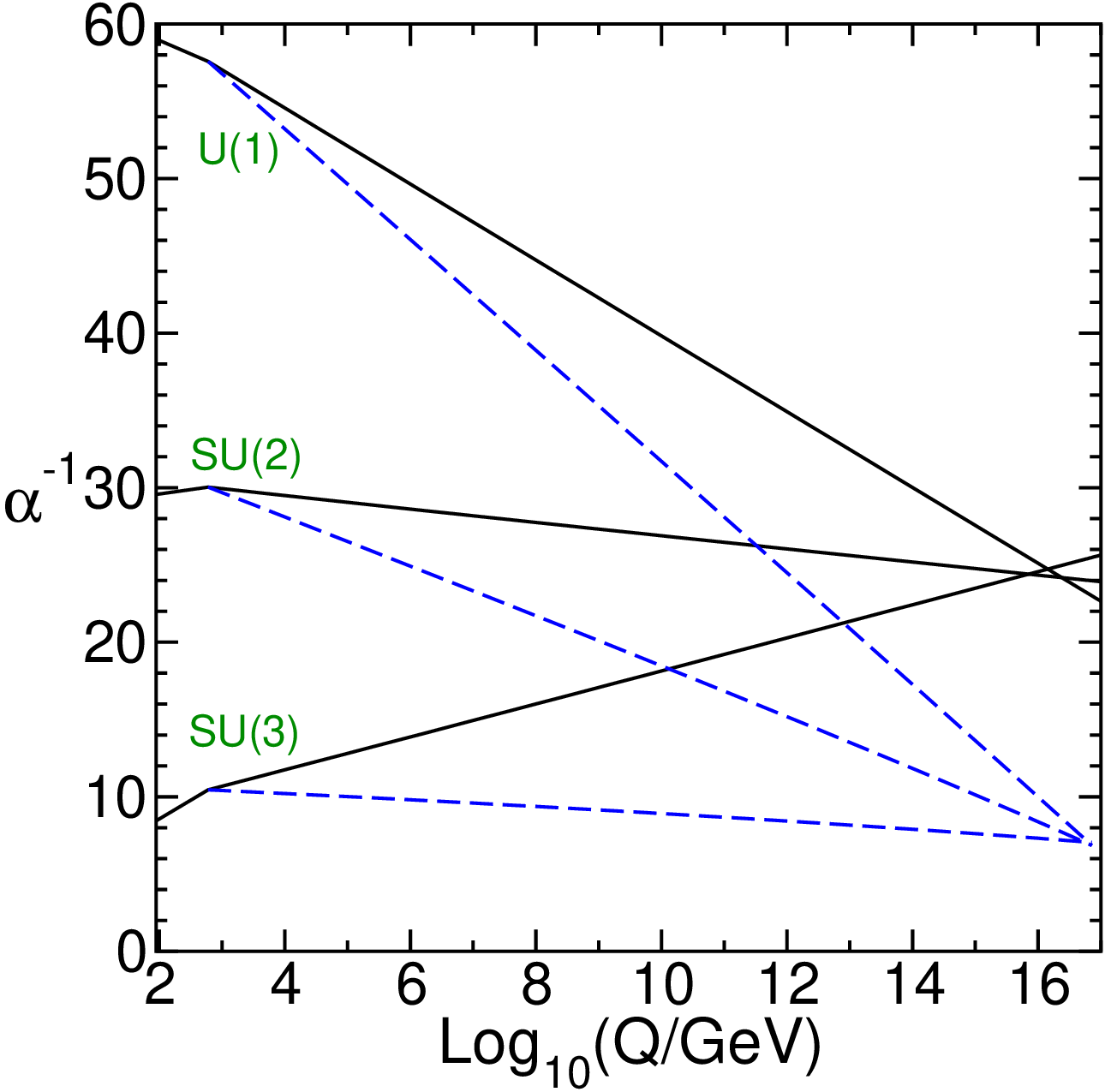}
\end{flushleft}
\end{minipage}
\begin{minipage}[]{0.47\linewidth}
\caption{\label{fig:unification}
Gauge coupling unification in the MSSM (solid lines) and in the OTLEE 
model of eq.~(\ref{OTLEE})  
(dashed blue lines). The running is performed with 2-loop beta functions, 
with all particles beyond the Standard Model taken to decouple at $Q = 
600$ GeV, and $m_t = 173.1$ GeV with $\tan\!\beta = 10$,
with the extra Yukawa couplings set to 0 for simplicity.
}
\end{minipage}
\end{figure}

In contrast, the OTLEE, TLUDD, and TLEDDD models all allow\footnote{The 
OTLEE and TLEDDD models can also have Yukawa couplings $H_u \overline L 
E$, $H_d L \overline E$ (and $H_u L \overline N$ and $H_d \overline L 
N$ if singlets are present), but I will assume these are absent or 
negligible for simplicity. If present, they would reduce the quasi-fixed 
point values of $\yk, \yh$.} the qualitatively new possibility of 
(doublet)-(triplet)-(doublet) superpotential Yukawa couplings $\yk$ and 
$\yh$ involving the MSSM Higgs fields $H_u, H_d$ and the weak isotriplet 
$T$ field and the new vector-like isodoublet fields $L$ and $\overline 
L$. Including also the relevant gauge-singlet mass terms, the 
superpotential is:
\beq
W = 
\yk H_u T L + 
\yh H_d T \overline L +
\frac{1}{2} M_T T^2 + M_L \overline L L .
\label{eq:WTLLsimple}
\eeq
In this paper, I will examine the features of models that include this 
structure. In particular, when $\yk$ is large, it can induce a 
significant positive correction to $m_{h^0}$. The infrared quasi-fixed 
point for $\yk$ is not too small to do so, in part because of the larger 
$SU(2)_L$ Casimir invariant for the triplet $T$ compared to a doublet (2 
compared to $3/4$). In the following, I will use the OTLEE model as an 
example, but many of the results apply also to the TLUDD and TLEDDD 
models with only small numerical changes. The unification of the gauge 
couplings in the OTLEE model is shown in Figure \ref{fig:unification}, 
with $\yk=\yh = 0$ for simplicity.
Although the $SU(3)_c$ gauge coupling would not run according to the 
one-loop renormalization group (RG) equations, two-loop effects are 
seen to cause it to get stronger in the ultraviolet, but not enough to 
become non-perturbative before unification takes place. The runnings in 
the TLUDD and TLEDDD models are only slightly different; all three of 
these models have $N=3$ from eq.~(\ref{eq:defN}).

\section{The new particles and their masses}
\label{sec:content}
\setcounter{equation}{0}
\setcounter{footnote}{1}

In this section, I consider the fermion and scalar content of the $T = 
(T^+, T^0, T^-)$, $L = (L^0, L^-)$, and $\overline L = (\overline L^+, 
\overline L^0)$ supermultiplets. After the mixing implied by the Yukawa 
couplings $\yk$ and $\yh$ in eq.~(\ref{eq:WTLLsimple}), the fermions will 
consist of three neutral Majorana fermion mass eigenstates, and two 
charged Dirac mass eigenstates denoted here by $\psi_i^0$ for $i=1,2,3$, 
and $\psi^\pm_i$ for $i=1,2$, respectively. To find the mass eigenstates 
and their mixing angles, the superpotential eq.~(\ref{eq:WTLLsimple}) can 
be written explicitly in terms of the different electric charge 
components of the gauge eigenstate fields as:
\beq
W &=& 
M_T (T^+ T^- + \frac{1}{2} T^0 T^0)
+ M_L ( L^- \overline L^+ - L^0 \overline L^0)
\nonumber \\ &&
+ \yk (T^0 L^- H_u^+ + T^0 L^0 H_u^0 + \sqrt{2} T^+ L^- H_u^0 
- \sqrt{2} T^- L^0 H_u^+)
\nonumber \\ &&
+ \yh (T^0 \overline L^+ H_d^- + T^0 \overline L^0 H_d^0 
+ \sqrt{2} T^+ \overline L^0 H_d^- 
       - \sqrt{2} T^- \overline L^+ H_d^0)
.
\eeq
Therefore, the mass matrices after electroweak symmetry 
breaking are, in two-component fermion notation \cite{DHM}:
\beq
{\cal L} &=& -\frac{1}{2} 
             \begin{pmatrix}T^0 & L^0 & \overline L^0 \end{pmatrix} {\cal M}_0
           \begin{pmatrix} T^0 \cr L^0 \cr \overline L^0 \end{pmatrix} - 
\begin{pmatrix} T^- & L^- \end{pmatrix} {\cal M}_\pm 
\begin{pmatrix} T^+ \cr \overline{L}^+ \end{pmatrix}
+ {\rm c.c.}
,
\\
{\cal M}_0 &=& 
\begin{pmatrix}
M_T & \yk v_u & \yh v_d \cr
\yk v_u & 0 & -M_L \cr
\yh v_d & -M_L & 0
\end{pmatrix}
, 
\qquad\>\>
{\cal M}_\pm \,=\, 
\begin{pmatrix}
M_T & -\sqrt{2} \yh v_d\cr 
\sqrt{2} \yk v_u & M_L
\end{pmatrix}
,
\label{eq:defMferms}
\eeq
where $v_u$ and $v_d$ are the VEVs of the
Higgs fields $H_u^0$ and $H_d^0$, with $v_u/v_d = \tan\!\beta$.
The real positive fermion mass eigenvalues 
and unitary mixing matrices $\Nzero$, $\Nminus$, and 
$\Nplus$ are defined by
\beq
\Nzero^* {\cal M}_0 \Nzero^\dagger &=& 
{\rm diag}(m_{\psi^0_1}, m_{\psi^0_2}, m_{\psi^0_3})
,
\label{eq:defineN}
\\
\Nminus^* {\cal M}_\pm \Nplus^\dagger &=&
{\rm diag}(m_{\psi^+_1}, m_{\psi^+_2}),
\label{eq:defineUV}
\eeq
with $(T^0, L^0, \overline L^0)_j = N_{ij}^* \psi^0_i$ and
$(T^-, L^-)_j = U_{ij}^* \psi^-_i$ and
$(T^+, \overline{L}^+)_j = V_{ij}^* \psi^+_i$.

The scalar components of the $T$, $L$, $\overline L$ supermultiplets
mix to form four complex charged scalars $\phi_i^\pm$ for $i=1,\ldots,4$,
and six real neutral scalars $\phi_i^0$ for $i=1,\ldots ,6$. 
The general form of the soft supersymmetry-breaking Lagrangian is:
\beq
-{\cal L}_{\rm soft} 
&=& 
a_{\yk} (T^0 L^- H_u^+ + T^0 L^0 H_u^0 + \sqrt{2} T^+ L^- H_u^0 
- \sqrt{2} T^- L^0 H_u^+)
\nonumber \\ &&
+ a_{\yh} (T^0 \overline L^+ H_d^- + T^0 \overline L^0 H_d^0 
+ \sqrt{2} T^+ \overline L^0 H_d^- 
- \sqrt{2} T^- \overline L^+ H_d^0)
\nonumber \\ &&
+ b_T (T^+ T^- + \frac{1}{2} T^0 T^0)
+ b_L ( L^- \overline L^+ - L^0 \overline L^0)
+ {\rm c.c.}
\nonumber \\ &&
+ m_T^2 (|T^0|^2 + |T^+|^2 + |T^-|^2) +
m_L^2 (|L^0|^2 + |L^-|^2) + 
m_{\overline L}^2 (|\overline L^+|^2 + |\overline L^0|^2) .
\eeq
It follows that the $6\times 6$ gauge-eigenstate 
squared-mass matrix for the neutral scalars is
\beq
\begin{pmatrix} C & D^\dagger \cr
                D & C
\end{pmatrix}
,
\label{eq:m2neutralscalars}
\eeq
in $3\times 3$ blocks, where
\beq
C &=& 
{\cal M}_0^\dagger {\cal M}_0 + 
{\rm diag} (m_T^2, \> m_L^2 + \Delta_{\frac{1}{2},0}, \>
m_{\overline L}^2 + \Delta_{-\frac{1}{2},0} )
\eeq
with electroweak $D$-term contributions defined by $\Delta_{T_3,q} = 
\frac{1}{2} [
T_3 g^2 + (T_3 - q) g^{\prime 2}
](v_d^2 - v_u^2) 
$, 
and
\beq
D = 
\begin{pmatrix}
b_T & a_{\yk} v_u - \yk \mu^* v_d & a_{\yh} v_d - \yh \mu^* v_u \cr
a_{\yk} v_u - \yk \mu^* v_d & 0 & -b_L \cr
a_{\yh} v_d - \yh \mu^* v_u & -b_L & 0
\end{pmatrix}
.
\eeq
For the charged scalars the $4\times 4$
gauge-eigenstate squared-mass matrix is:
\beq
\begin{pmatrix} E & G^\dagger \cr
                G & F
\end{pmatrix}
,
\label{eq:m2chargedscalars}
\eeq
where the $2\times 2$ blocks are
\beq
E &=& {\cal M}_\pm^\dagger {\cal M}_\pm + 
{\rm diag} (m_T^2 + \Delta_{1,1}, \> 
m_{\overline L}^2 + \Delta_{\frac{1}{2},1})
,
\\
F &=& {\cal M}_\pm {\cal M}_\pm^\dagger + 
{\rm diag} (m_T^2 + \Delta_{-1,-1}, \> 
m_{L}^2 + \Delta_{-\frac{1}{2},-1})
,
\\
G &=& \begin{pmatrix}
b_T & \sqrt{2}(-a_{\yh} v_d + \yh \mu^* v_u) \cr
\sqrt{2} (a_{\yk} v_u - \yk \mu^* v_d) & b_L
\end{pmatrix}
.
\label{eq:defmatG}
\eeq
The tree-level scalar squared masses $m^2_{\phi^0_i}$ and 
$m^2_{\phi^\pm_i}$ are the eigenvalues of 
eqs.~(\ref{eq:m2neutralscalars}) and (\ref{eq:m2chargedscalars}). I will 
assume that, as usual in phenomenologically viable supersymmetric models, 
the soft terms $m_T^2$, $m_L^2$, and $m_{\overline L}^2$ are large enough 
to make the scalar mass eigenstates $\phi^0_i$ and $\phi^\pm_i$ much 
heavier than their fermion counterparts $\psi^0_i$ and $\psi^\pm_i$.

\begin{figure}[!tp]
\begin{minipage}[]{0.49\linewidth}
\begin{flushright}
\includegraphics[width=6.7cm,angle=0]{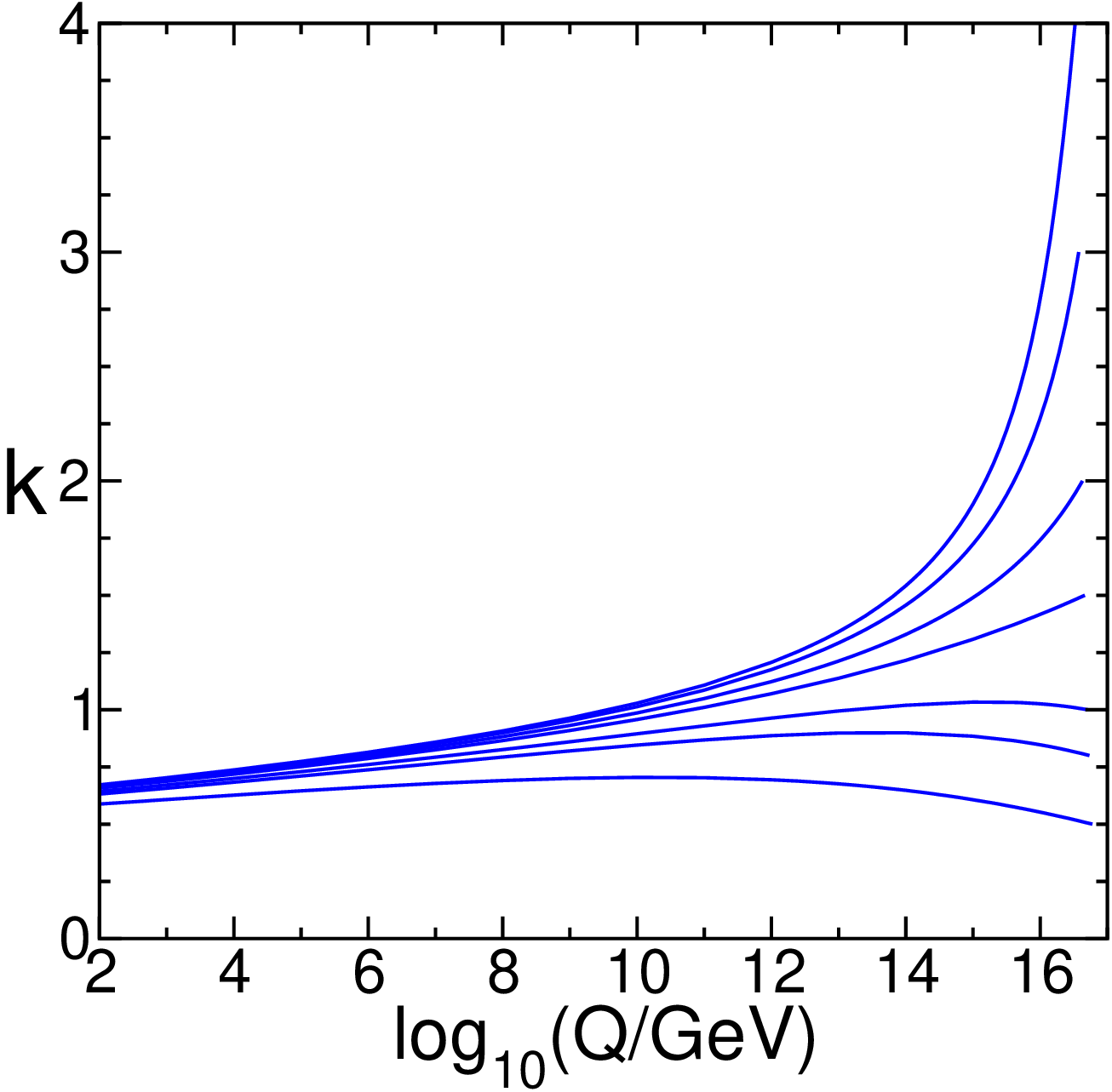}
\end{flushright}
\end{minipage}
~~~~~~\begin{minipage}[]{0.4\linewidth}
\caption{\label{fig:krun}
The running of the Yukawa coupling $\yk$
for various 
different input
values at the unification scale,
with $\yh = 0$ and $\tan\beta = 10$. This illustrates the quasi-fixed point 
structure, leading to $\yk \approx 0.69$ at $Q = 500$ GeV.}
\end{minipage}
\end{figure}

\begin{figure}[!tp]
\begin{minipage}[]{0.49\linewidth}
\begin{flushright}
\includegraphics[width=6.9cm,angle=0]{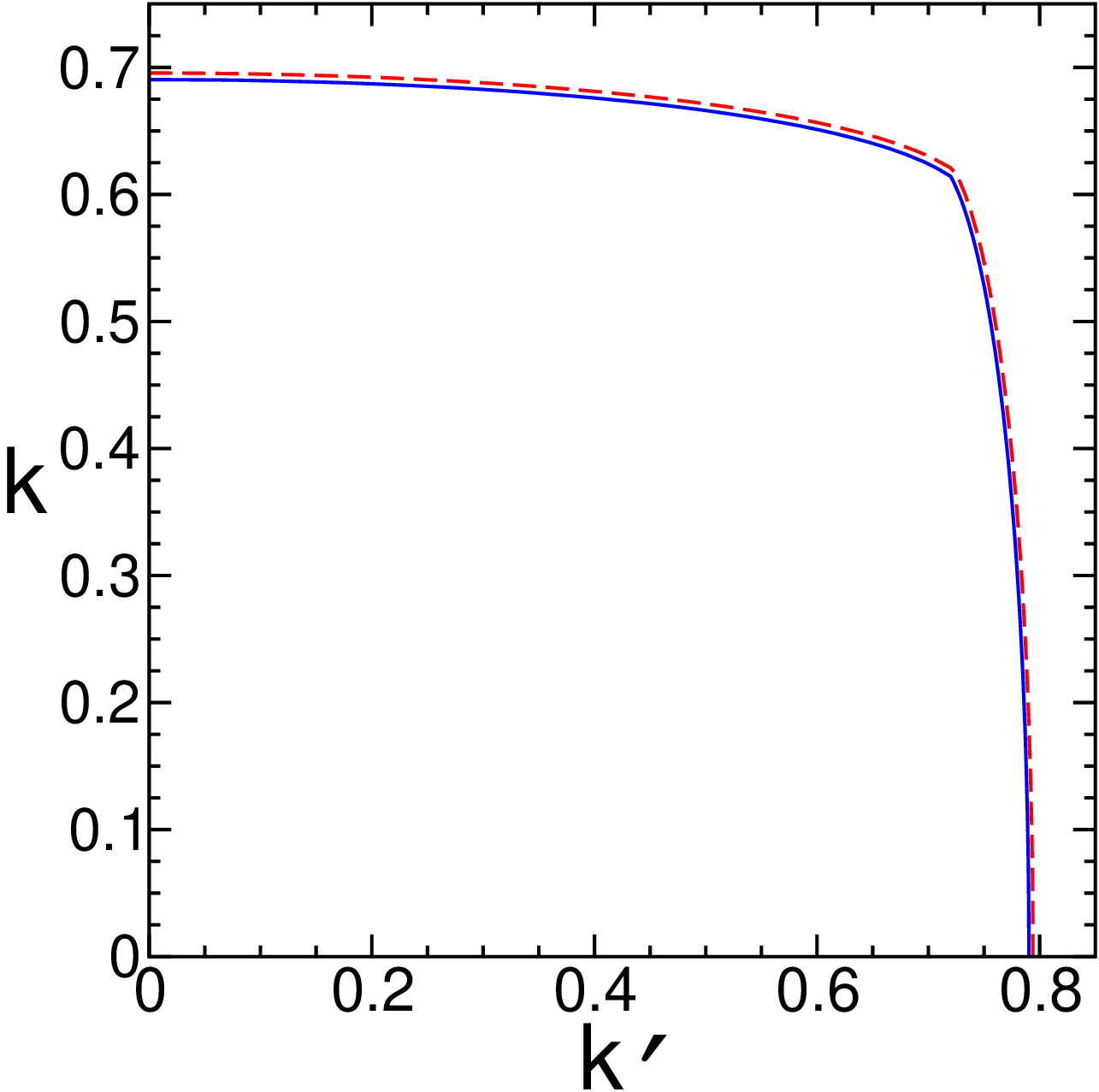}
\end{flushright}
\end{minipage}
~~~~~~\begin{minipage}[]{0.4\linewidth}
\caption{\label{fig:hkcon}
The infrared quasi-fixed point contours for the Yukawa couplings $\yk$ 
and $\yh$ evaluated at a renormalization scale of $Q=500$ GeV (solid line)
and $Q = 1000$ GeV (dashed line), with $\tan\beta = 10$. The regions below and to the left of 
the contours are allowed by $\yk, \yh < 3$ at the unification scale.}
\end{minipage}
\end{figure}

An important feature of these models is that infrared quasi-fixed points \cite{PRH}
govern the new Yukawa couplings. This can be seen 
qualitatively from the one-loop parts of the RG equations:
\beq
Q\frac{d\yk}{dQ} = \beta_{\yk} &=& \frac{\yk}{16 \pi^2} \bigl (
8 \yk^2 + 2 \yh^2 + 3 y_t^2 - 7 g_2^2 -\frac{3}{5} g_1^2 
\bigr ) 
,
\\
Q\frac{d\yh}{dQ} =\beta_{\yh} &=& \frac{\yh}{16 \pi^2} \bigl (
8 \yh^2 + 2 \yk^2 + 3 y_b^2 + y_\tau^2 - 7 g_2^2 -\frac{3}{5} g_1^2 
\bigr ) 
.
\eeq
The infrared quasi-fixed points occur when the positive contributions 
from Yukawa couplings nearly cancel the negative contributions from gauge 
couplings. In the following, we will be most interested in the case that 
$\yk$ is a large as possible, because when $\tan\!\beta > 1$ this leads 
to the largest possible contribution to the mass of $h^0$; this is 
obtained when $\yh = 0$. The two-loop RG running of $\yk$ for various 
different input values at the unification scale is shown in Figure 
\ref{fig:krun}.
More generally, the contour of quasi-fixed points in the $(\yh, \yk)$
plane is shown in Figure \ref{fig:hkcon}, obtained by requiring
the perturbativity condition\footnote{This criterion is somewhat arbitrary, but the fixed point values are not very sensitive to it.} 
$\yk,\yh < 3$ at the unification scale. 
Although there is coupling between $\yk$ and $\yh$ in their RG equations, 
the quasi-fixed point value of $\yk$ does not vary much as long as $\yh$ 
is not too large. In the following, I will use $\yk = 0.69$ as the fixed 
point value, motivated by the fact that a wide range of input values at 
the unification scale will end up close to this fixed point.

\begin{figure}[!tp]
\begin{minipage}[]{0.49\linewidth}
\begin{flushleft}
\includegraphics[width=7.2cm,angle=0]{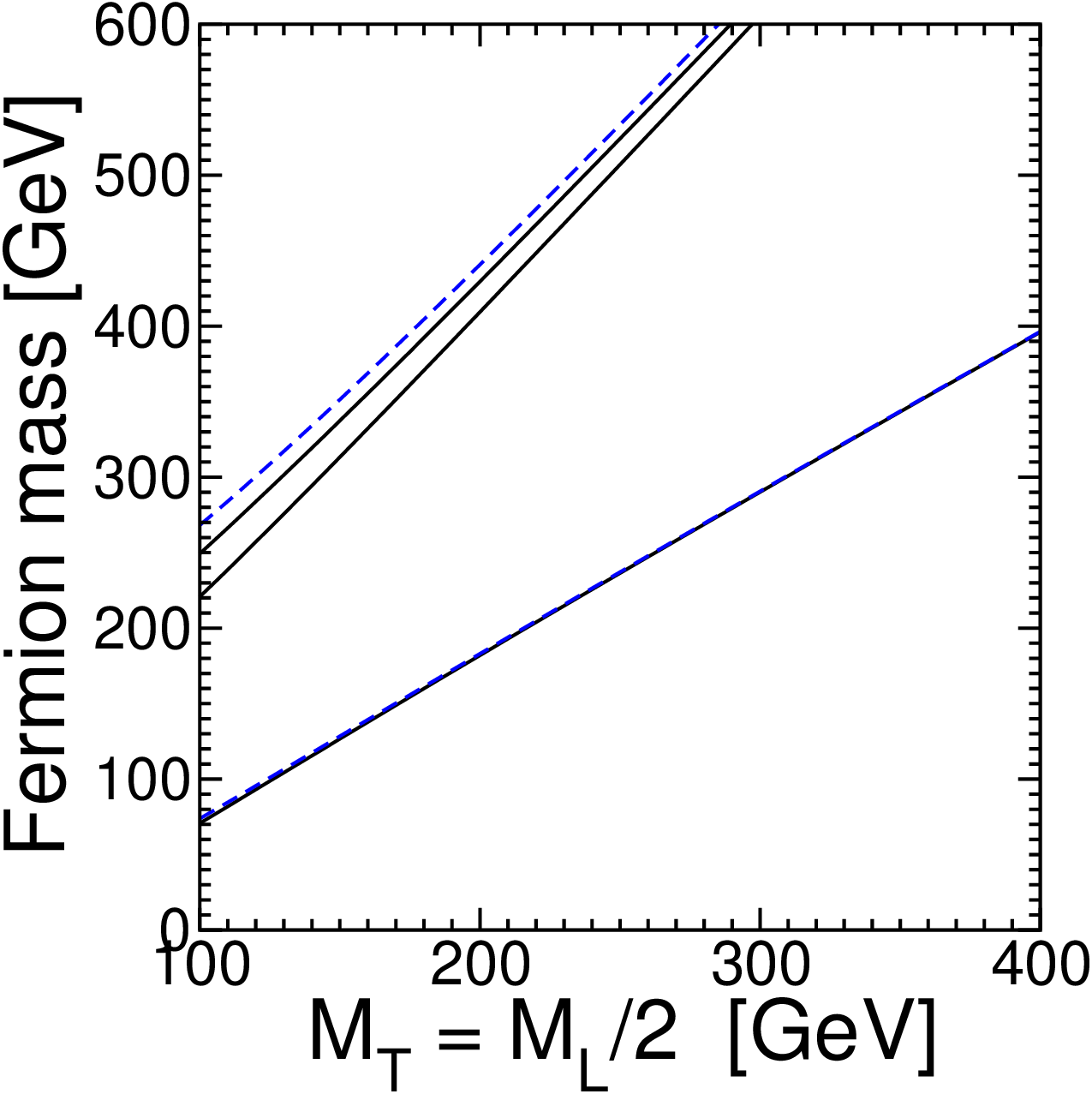}
\end{flushleft}
\end{minipage}
\begin{minipage}[]{0.49\linewidth}
\begin{flushright}
\includegraphics[width=7.2cm,angle=0]{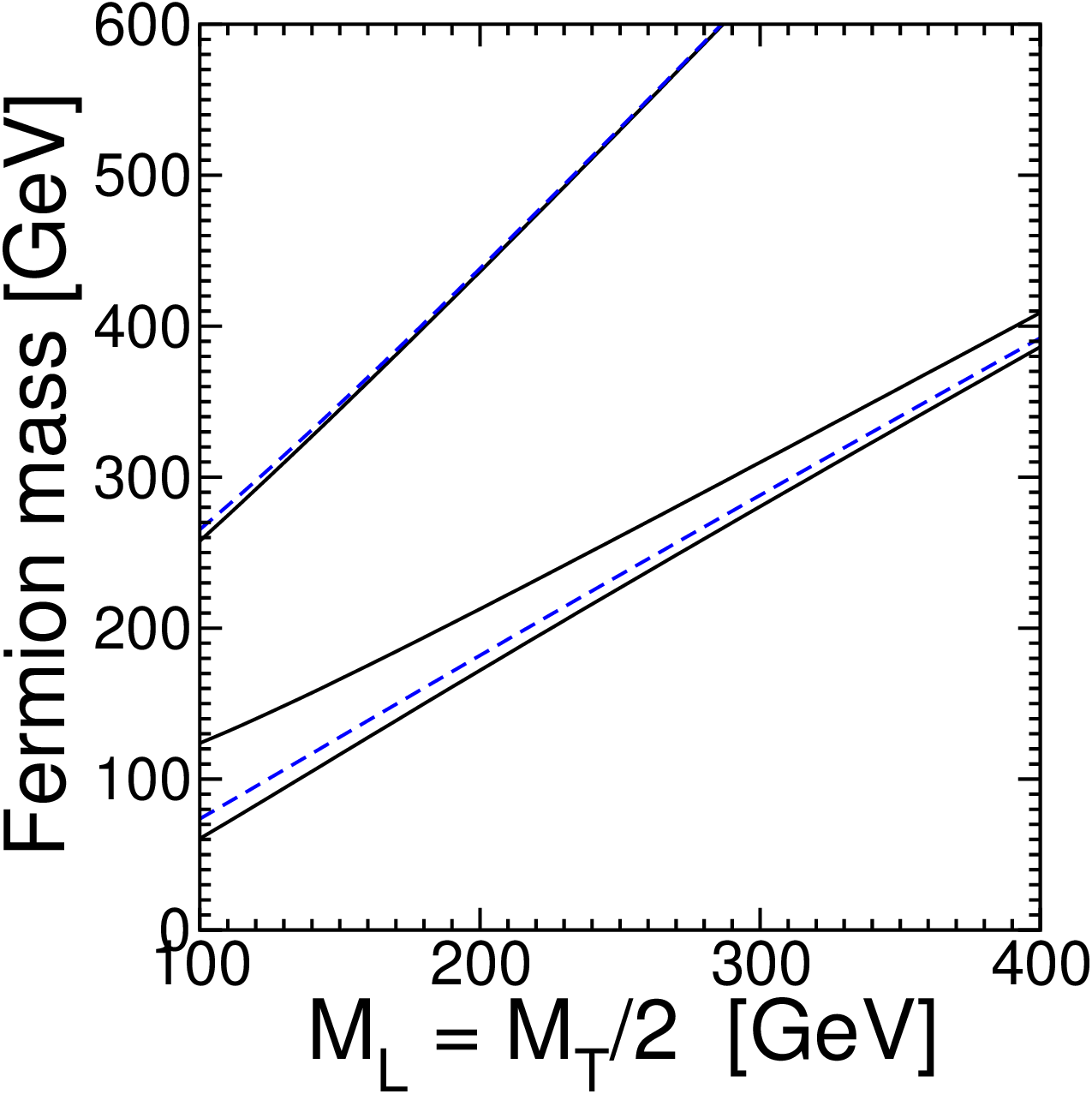}
\end{flushright}
\end{minipage}

\vspace{0.1cm}

\begin{minipage}[]{0.49\linewidth}
\begin{flushleft}
\includegraphics[width=7.2cm,angle=0]{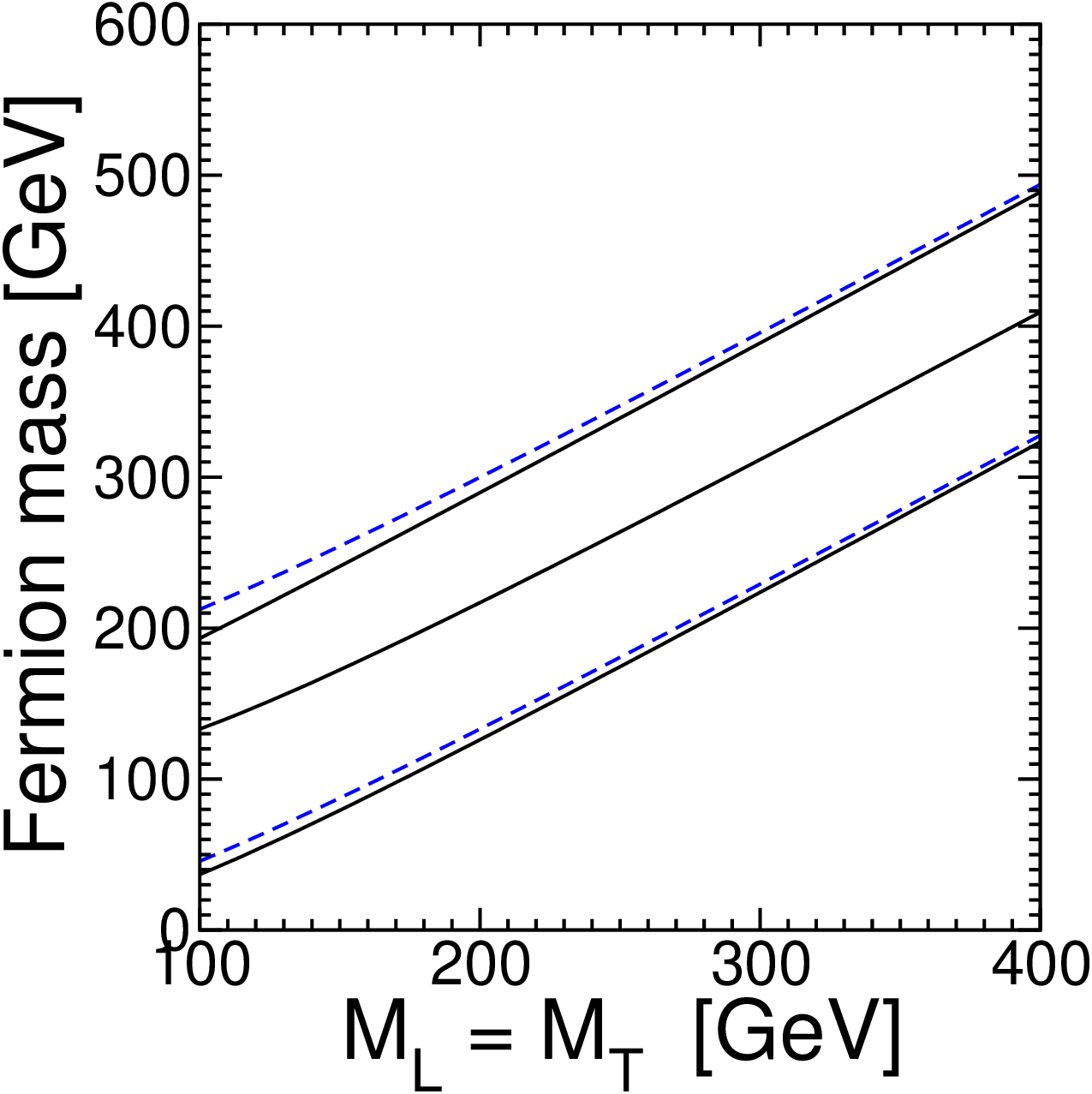}
\end{flushleft}
\end{minipage}
\begin{minipage}[]{0.49\linewidth}
\caption{\label{fig:fermionmasses}
The spectrum of neutral (solid lines) and charge $\pm 1$ (dashed lines)
fermion masses, as a function of the superpotential mass parameters
$M_T = M_L/2$ (upper left),
$M_L = M_T/2$ (upper right),
and 
$M_L = M_T$ (lower left). 
The lighter fermions in these three cases are respectively mostly
triplet, mostly doublet, and mixed triplet-doublet.
Here $\yk = 0.69$ and $\yh=0$ are assumed.}
\end{minipage}
\end{figure}

The phenomenology of these models will depend strongly on the 
fermion masses. These masses are shown in Figure \ref{fig:fermionmasses}
for $(\yk, \yh) = (0.69, 0)$ and $\tan\!\beta = 10$ and 
varying superpotential mass parameters
$M_T$ and $M_L$, for three different fixed ratios $M_T/M_L = 0.5$, 1, and 2.
One-loop radiative corrections to the masses are potentially important,
and so are included using the results of \DeltaMappendix.
In all cases, the lightest of the new
fermions turns out to be the neutral $\psi^0_1$.

\begin{figure}[!tp]
\begin{minipage}[]{0.49\linewidth}
\begin{flushleft}
\includegraphics[width=7.5cm,angle=0]{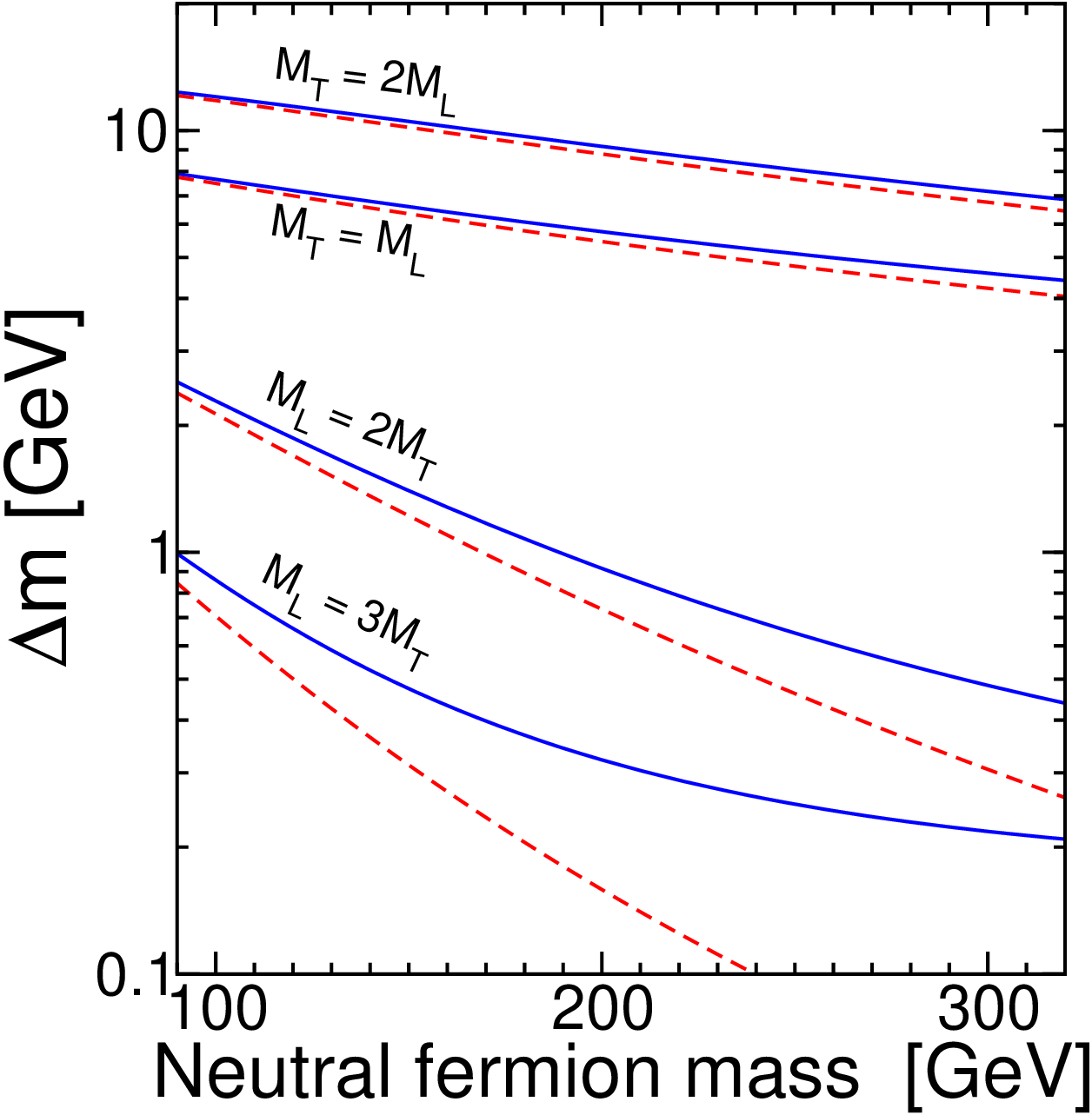}
\end{flushleft}
\end{minipage}
\begin{minipage}[]{0.49\linewidth}
\caption{\label{fig:deltafermionmasses}
The mass difference $\Delta m = m_{\psi_1^\pm} - 
m_{\psi_1^0}$ between the next-lightest (charged) and lightest (neutral) 
new fermion masses, as a function of $m_{\psi_1^0}$. 
The four solid lines correspond to 
the cases, from top to bottom, $M_T = 2M_L$, $M_T = M_L$, $M_L = 2 M_T$,
and $M_L = 3 M_T$, including one-loop radiative corrections.
For comparison, the dashed lines show what the results would be with
the one-loop corrections omitted.
Here $\yk=0.69$ and $\yh = 0$ and $\tan\!\beta = 10$ are assumed.
Note that $\Delta m$ is always positive, and decreases 
as the lightest fermions become more triplet-like,
but is prevented from becoming too small by the radiative corrections.}
\end{minipage}
\end{figure}

When $M_T < M_L$, the lightest fermions $\psi_1^0$ and $\psi_1^\pm$ form 
a very nearly degenerate triplet, but the presence of the Yukawa coupling 
$\yk$ and one-loop radiative corrections ensures a non-zero splitting. 
When $M_L < M_T$, the lightest fermions $\psi_1^0$, $\psi_2^0$, 
$\psi_1^\pm$ are mostly a Dirac pair of doublets, with a much larger mass 
splitting than the light triplet case. When $M_L \sim M_T$, 
there is significant mixing between the doublets and the triplet, 
although the splitting between $m_{\psi_1^\pm}$ and $m_{\psi_1^0}$ can be 
seen to remain fairly small. The mass splitting between the lowest-lying 
states
\beq
\Delta m \equiv m_{\psi_1^\pm} - m_{\psi_1^0}
\eeq
plays an important role in collider signals, and so is shown in
Figure \ref{fig:deltafermionmasses} for cases with the
lightest fermions mostly doublets $(M_T = 2 M_L)$, mixed $(M_T = M_L)$,
and mostly triplet ($M_L = 2 M_T$ and $M_L = 3 M_T$).
The one-loop radiative corrections
always increase $\Delta m$. The mass splitting is smallest in 
the extreme limit of pure winos ($\yk v_u \ll M_T \ll M_L$) where it
asymptotically approaches $\Delta m = 0.16$ GeV
\cite{Cheng:1998hc,Feng:1999fu,Gherghetta:1999sw}.
However, in most cases the mass splitting is considerably larger
because of the Yukawa coupling, 
and it is always larger than the charged pion mass.

The RG running of the soft supersymmetry breaking terms has several 
interesting features that are comparable to those found in the LND, QUE 
and QDEE models studied in \cite{Martin:2009bg}. [To be concrete, here I 
use $\tan\!\beta = 10$, $m_t = 173.1$, and  
$(k,h)=(3,0)$ at the unification scale. It cannot be under-emphasized that 
working to only one-loop order would yield very misleading results, 
because of the large values of the gauge couplings and non-trivial 
running of $g_3$ at high scales.] 
First, if one 
assumes that the gaugino masses are unified with a value $m_{1/2}$ at the 
same scale as the gauge couplings, then one finds that the RG 
running leads to quite different ratios than in the MSSM,
\beq
(M_1, M_2, M_3) &=&  (0.13,\, 0.23,\, 0.47) m_{1/2},\qquad \mbox{(OTLEE model)}
\label{eq:MgauginosOTLEE}
\\
&=&  (0.13,\, 0.24,\, 0.62) m_{1/2},\qquad \mbox{(TLUDD, TLEDDD models)}
\label{eq:MgauginosTLUDD}
\\
&=&  (0.41,\, 0.77,\, 2.28) m_{1/2},\qquad \mbox{(MSSM)},
\label{eq:MgauginosMSSM}
\eeq
evaluated at $Q = 1$ TeV.
In particular, the ratios $M_3/M_2$ and $M_3/M_1$ are both much smaller 
in the extended models than in the MSSM. The extended models therefore 
predict a more compressed spectrum of 
superpartners than is found in the MSSM with unified gaugino masses.

If one takes the soft scalar squared masses and (scalar)$^3$ terms to 
vanish at the unification scale, corresponding to the ``no-scale" or 
``gaugino-mediated" boundary conditions $m_0^2=0$ and $A_0 = 0$, then one 
finds for the ordinary first- and second-family squark and slepton mass 
parameters at $Q = 1$ TeV:
\beq
&&(m_{\widetilde q_1}, m_{\widetilde u_1}, m_{\widetilde d_1}, 
 m_{\widetilde \ell_1}, m_{\widetilde e_1}) 
\nonumber 
\\ &=& (1.15,\, 1.08,\, 1.08,\, 0.50,\, 0.30) m_{1/2} 
\qquad \mbox{(OTLEE model)},
\label{eq:OTLEEMSSMscalars}
\\
&=& (1.29,\, 1.23,\, 1.22,\, 0.51,\, 0.30\,) m_{1/2} 
\qquad \mbox{(TLUDD, TLEDDD models)},
\\
&=& (2.08,\, 2.01,\, 2.00,\, 0.67,\, 0.37) m_{1/2} 
\qquad \mbox{(MSSM)}.
\eeq
Again, one sees a compression of the mass spectrum for the extended models
compared to the MSSM.
The soft masses for the new scalars in the OTLEE, TLUDD, and 
TLEDDD models are, respectively:
\beq
(m_{\widetilde T}, m_{\widetilde L}, m_{\widetilde{\overline{L}}}, 
m_{\widetilde E}, m_{\widetilde{\overline E}}, m_{\widetilde O}) 
&=& (0.73,\, 0.29,\, 0.51,\, 0.29,\, 0.30,\, 1.51) m_{1/2} ,
\label{eq:OTLEEmTmL}
\\
(m_{\widetilde T}, m_{\widetilde L}, m_{\widetilde{\overline{L}}}, 
m_{\widetilde U}, m_{\widetilde{\overline U}}, 
m_{\widetilde D}, m_{\widetilde{\overline D}}) 
&=& (0.74,\, 0.33,\, 0.51,\, 1.23,\, 1.23,\, 1.22,\, 1.22) m_{1/2} ,
\\
(m_{\widetilde T}, m_{\widetilde L}, m_{\widetilde{\overline{L}}}, 
m_{\widetilde E}, m_{\widetilde{\overline E}}, 
m_{\widetilde D}, m_{\widetilde{\overline D}}) 
&=& (0.75,\, 0.33,\, 0.51,\, 0.29,\, 0.30,\, 1.23,\, 1.23) m_{1/2} .
\label{eq:TLEDDDnewscalars}
\eeq
Comparing eqs.~(\ref{eq:OTLEEMSSMscalars})-(\ref{eq:TLEDDDnewscalars})
to eqs.~(\ref{eq:MgauginosOTLEE})-(\ref{eq:MgauginosMSSM}) 
shows that, unlike the MSSM, the extended models considered here permit 
gaugino mass domination for the soft supersymmetry breaking terms at the unification scale while 
still having a bino-like neutralino as the LSP.
(This feature was also observed in the QUE and QDEE models in 
ref.~\cite{Martin:2009bg}.)

\begin{figure}[!tp]
\begin{minipage}[]{0.49\linewidth}
\begin{flushright}
\includegraphics[width=7.5cm,angle=0]{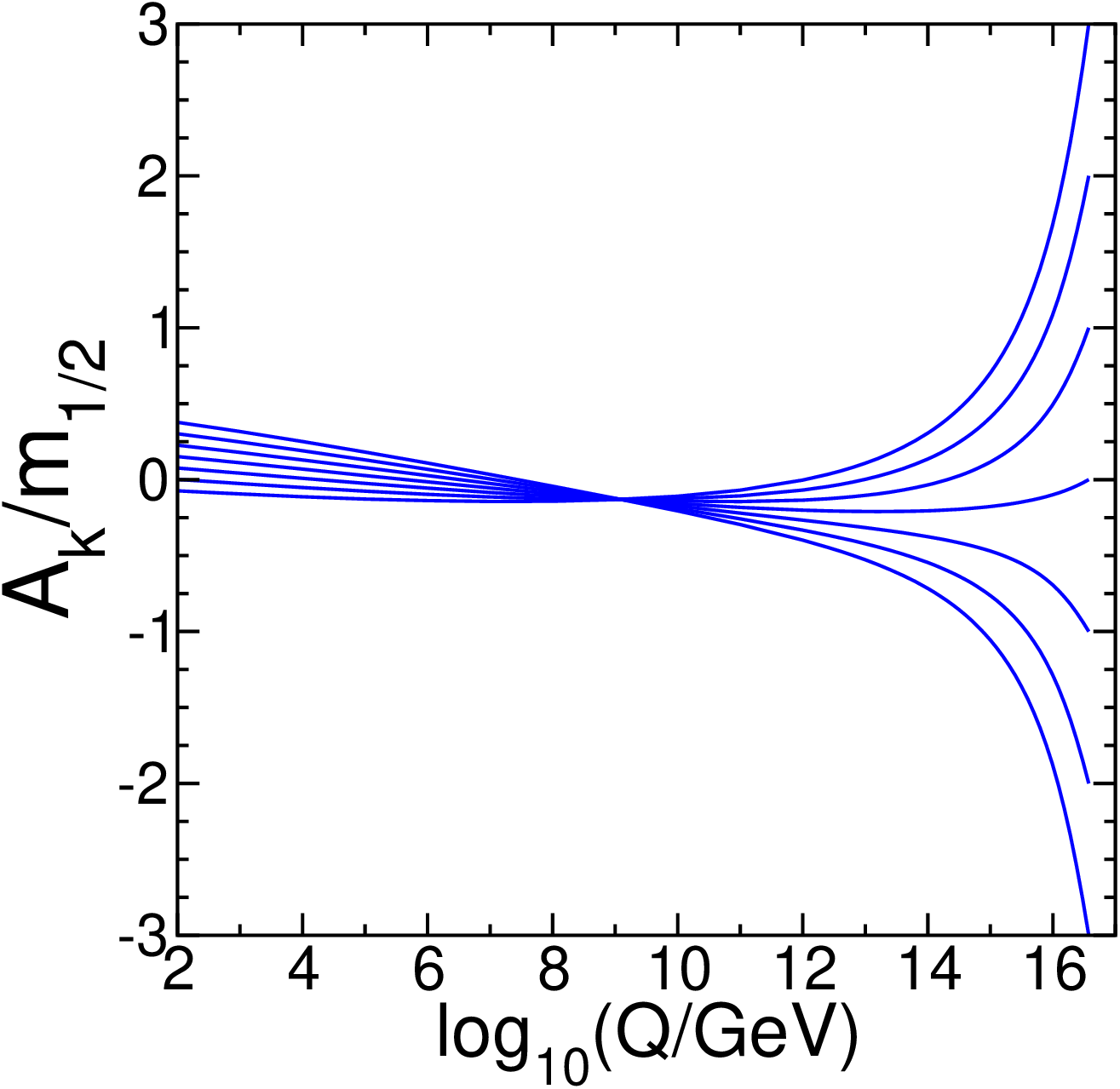}
\end{flushright}
\end{minipage}
~~~~~~
\begin{minipage}[]{0.4\linewidth}
\caption{\label{fig:Akfixed}
Renormalization group running of (scalar)$^3$ coupling parameter $A_{\yk} = 
a_{\yk}/\yk$, normalized by $m_{1/2}$, in the OTLEE model with $\yk$ near 
its quasi-fixed point trajectory (defined by $\yk = 3$ at the gauge 
coupling unification scale). This shows a strongly attractive quasi-fixed 
point behavior for $A_{\yk}$. Very similar results obtain for the 
TLUDD and TLEDDD models.}
\end{minipage}
\end{figure}

Another important consideration is the running of the (scalar)$^3$ 
coupling $a_{\yk}$. 
The coupling $a_{\yk}$ will play an important role in the corrections to 
$m_{h^0}$ to be discussed below. It turns out that 
when $\yk$ is near its quasi-fixed point trajectory, then 
the quantity
\beq
A_{\yk} \equiv a_{\yk}/\yk
\eeq
itself has a strongly attractive quasi-fixed point near small multiples
of $m_{1/2}$, as shown in Figure \ref{fig:Akfixed} for the OTLEE model.
I have checked that the TLUDD and TLEDDD models give very similar results,
and that this behavior is not very sensitive to the assumption of gaugino mass 
unification, if $m_{1/2}$ is replaced by the value of $M_2$ at the 
unification scale.

\section{Corrections to the lightest Higgs scalar boson mass}
\label{sec:deltamh}
\setcounter{equation}{0}
\setcounter{footnote}{1}

In this section, I consider the contribution of the new doublet and 
triplet supermultiplets $L$, $\overline L$, and $T$ to the lightest Higgs 
scalar boson mass. The effective potential approximation provides a 
simple way to estimate this contribution, and is equivalent to neglecting 
non-zero external momentum effects in $h^0$ self-energy diagrams. (This is 
an accurate approximation since $m^2_{h^0} < 4 m_{\psi_1^0}^2$.) The 
one-loop contribution to the effective potential due to the particles 
in the $L$, $\overline L$, and $T$ supermultiplets is:
\beq
\Delta V = 
\sum_{i=1}^6 F(m^2_{\phi^0_i}) 
- 2 \sum_{i=1}^3 F(m^2_{\psi^0_i})
+ 2 \sum_{i=1}^4 F(m^2_{\phi^\pm_i})
- 4 \sum_{i=1}^2 F(m^2_{\psi^\pm_i}) .
\label{eq:DeltaV}
\eeq
Here $m^2_{\phi^0_i}$, $m^2_{\psi^0_i}$, 
$m^2_{\phi^\pm_i}$, and $m^2_{\psi^\pm_i}$
are the VEV-dependent tree-level squared-mass eigenvalues
from eqs.~(\ref{eq:defMferms})-(\ref{eq:defineUV}), 
(\ref{eq:m2neutralscalars}), and
(\ref{eq:m2chargedscalars}),
and
$
F(x) = x^2 [\ln(x/Q^2) - 3/2]/64 \pi^2,
$
with $Q$ the renormalization scale. I will assume the decoupling 
approximation that the neutral Higgs mixing angle (in the standard 
convention, described e.g.~in \cite{primer}) is $\alpha = \beta - 
\pi/2$, which is valid if $m_{A^0}^2 \gg m^2_{h^0}$. Then the correction 
to $m^2_{h^0}$ is
\beq
\Delta m_{h^0}^2 = 
\left \lbrace \frac{\sin^2\!\beta}{2} \Bigl [ 
\frac{\partial^2}{\partial v_u^2} 
- \frac{1}{v_u}\frac{\partial}{\partial v_u}\Bigr ]
+ \frac{\cos^2\!\beta}{2} \Bigl [ 
\frac{\partial^2}{\partial v_d^2} 
- \frac{1}{v_d}\frac{\partial}{\partial v_d}\Bigr ]
+ \sin\!\beta\cos\!\beta 
\frac{\partial^2}{\partial v_u \partial v_d} \right \rbrace \Delta V
.
\label{eq:Deltam2h}
\eeq
In the OTLEE, TLUDD, and TLEDDD models, the other new fields 
$(O,E,\overline{E}, U, \overline{U}, D, \overline{D})$
do not make 
a significant radiative 
contribution to the Higgs mass, as they do not have Yukawa 
couplings to $H_u$ and $H_d$.

\begin{figure}[!tp]
\begin{minipage}[]{0.49\linewidth}
\begin{flushleft}
\includegraphics[width=7.7cm,angle=0]{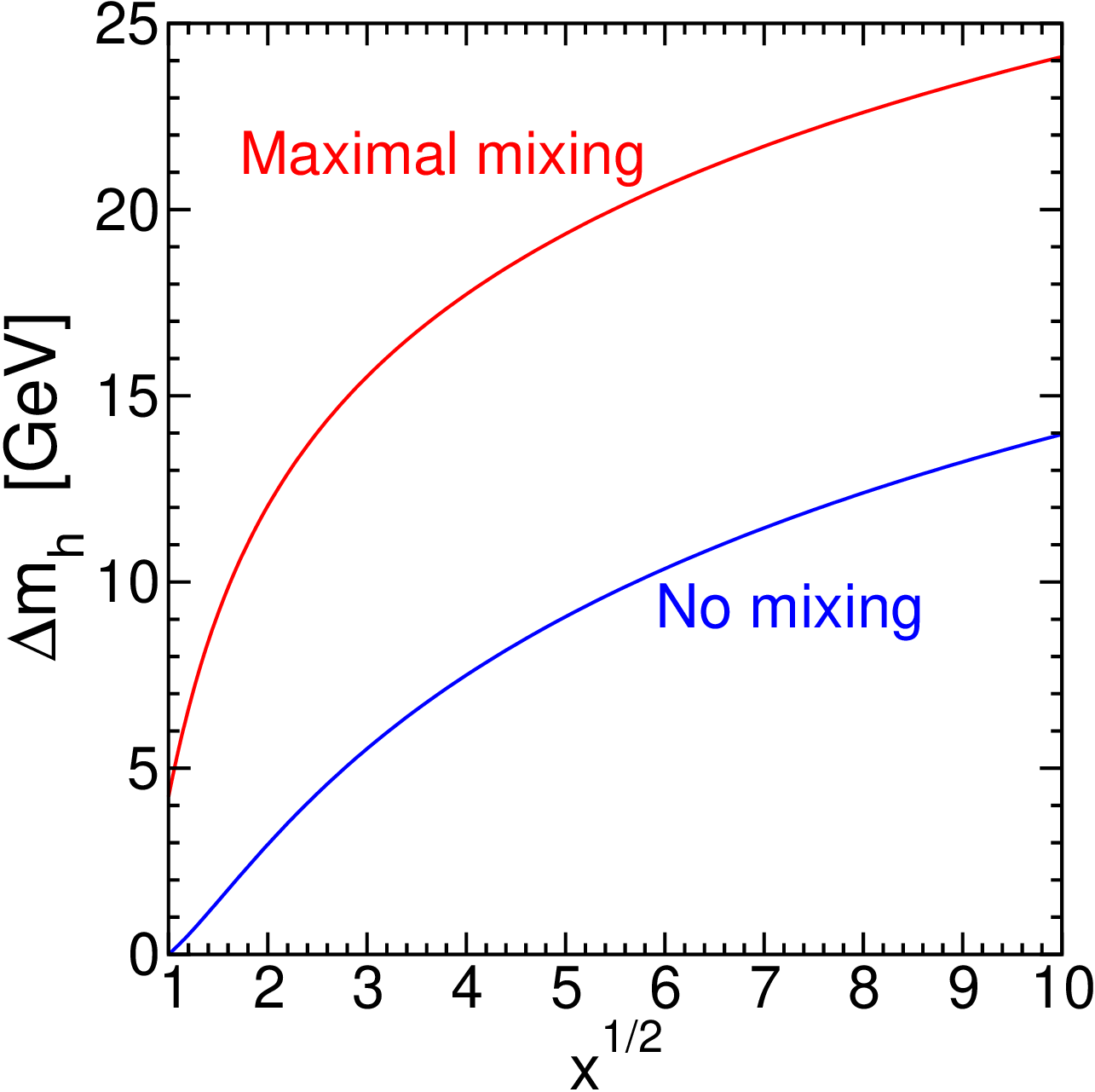}
\end{flushleft}
\end{minipage}
~~~~~
\begin{minipage}[]{0.4\linewidth}
\caption{\label{fig:deltamhsimple}
Estimates for the corrections to the $h^0$ mass as a function of 
$\sqrt{x} = M_S/M_F$, where $M_S$ and $M_F$ are the mean scalar and 
fermion masses in the supermultiplets $L$, $\overline L$, $T$, in the 
simplified model framework used in eq.~(\ref{eq:delmhsimple}) of the 
text, using $v^2 \yk^4\sin^4\!\beta = $ (83 GeV)$^2$, corresponding to the 
quasi-fixed point with reasonably large $\tan\!\beta$. The lower line is 
the no-mixing case $X_{\yk} = 0$, and the upper line is the ``maximal mixing" 
case $X^2_{\yk} = 6M_S^2 (1 - 2/5x).$ The $h^0$ mass before the correction 
is taken to be 110 GeV.}
\end{minipage}
\end{figure}

Before obtaining numerical results in a realistic model, 
it is useful to first consider a relatively simple analytical result
for the case that the superpotential mass parameters are equal
($M_T = M_L \equiv M_F$) and the non-holomorphic soft supersymmetry-breaking
squared masses are also equal 
($m_T^2 = m_L^2 = m_{\overline L}^2 \equiv m^2$),
and neglecting the holomorphic terms $b_T$ and $b_L$.
Then, writing
\beq
x &=& M_S^2/M_F^2,\qquad
M_S^2 \equiv M_F^2 + m^2 = \mbox{average scalar mass}  
\\
X_{\yk} &=& A_{\yk} - \mu \cot\!\beta,
\eeq
and, expanding in $\yk$, I find
\beq
\Delta m^2_{h^0} &=& \frac{v^2}{4 \pi^2} 
\biggl \lbrace \yk^4 \sin^4\!\beta \left [ f(x) 
+ \frac{X_{\yk}^2}{M_S^2} \left (5 - \frac{2}{x} \right )
- \frac{5 X_{\yk}^4}{12 M_S^4 } \right ]
\nonumber \\ &&
+ \frac{3}{4} (g^2 + g^{\prime 2}) \yk^2 \sin^2\!\beta \cos(2\beta)
\left [ \ln (M_S^2/Q^2) + X_{\yk}^2/2 M_S^2 \right ]
\biggr \rbrace
\label{eq:delmhsimple}
\eeq
where
\beq
f(x) = 5 \ln(x) 
-\frac{9}{2} + \frac{11}{2x} -\frac{1}{x^2}
.
\eeq
Note that $x$ is approximately the ratio of the mean squared masses of 
the scalars to the fermions and is therefore assumed greater than 1, 
while the mixing between the new triplet and 
the new doublet scalars is parameterized by $X_{\yk}$. Similar to the 
models discussed in 
\cite{Moroi:1991mg,Babu:2004xg,Babu:2008ge,Martin:2009bg,Graham:2009gy}, 
the contribution to $\Delta m_{h^0}^2$ does not decouple with the overall 
new particle mass scale, provided that there is a hierarchy maintained 
between the scalars and the fermions. 
The electroweak $D$-term contribution involving $g, g'$ is quite small,
provided one chooses a RG scale $Q \sim M_S$, and is neglected below.
The maximum possible contribution 
to $\Delta m^2_{h^0}$ occurs when $X^2_{\yk} = 6M_S^2 (1 - 2/5x)$, leading to 
a ``maximal mixing" result given by $\Delta m^2_{h^0} = \frac{v^2}{4 
\pi^2} \yk^4 \sin^4\!\beta f_{\rm max}(x)$, where \beq f_{\rm max}(x) = f(x) 
+ \frac{3}{5} (5 - 2/x)^2. \eeq In Figure \ref{fig:deltamhsimple},
I show an estimate of these corrections to $\Delta m_{h^0}$, using $v^2 
\yk^4\sin^4\!\beta = $ (83 GeV)$^2$, corresponding to the quasi-fixed 
point with reasonably large $\tan\!\beta$, and assuming that the 
predicted $h^0$ mass before the correction is 110 GeV, so that $\Delta 
m_{h^0} = \sqrt{\mbox{(110 GeV)}^2 + \Delta m_{h^0}^2} - $ 110 GeV.

\begin{figure}[!tp]
\begin{minipage}[]{0.49\linewidth}
\begin{flushright}
\includegraphics[width=7.6cm,angle=0]{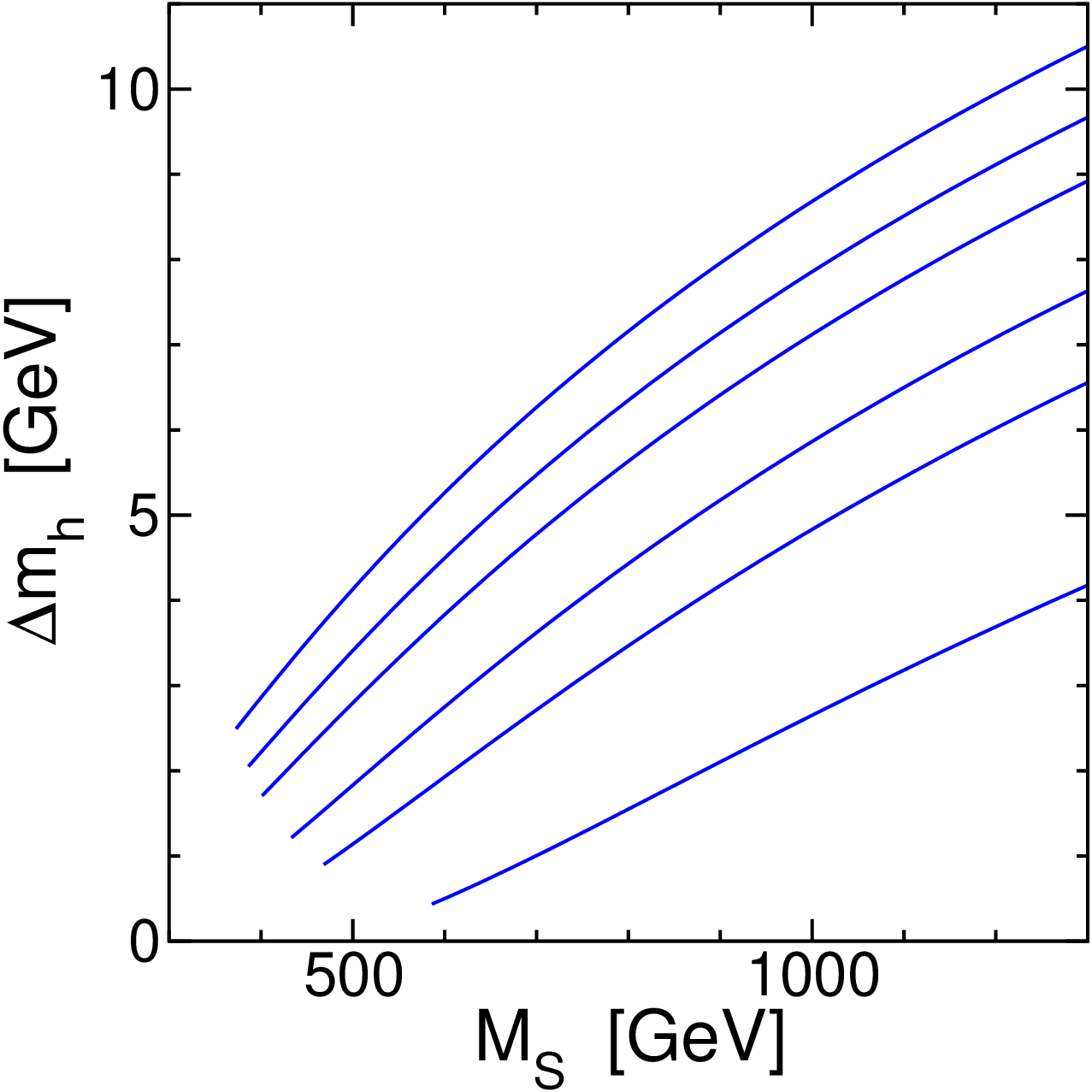}
\end{flushright}
\end{minipage}
~~~~~
\begin{minipage}[]{0.4\linewidth}
\caption{\label{fig:delmh}
Corrections to $m_{h^0}$, for $\yk = 0.69$ and $M_T = M_L \equiv M_F$ 
in the gaugino mass dominated scenario  
$(m_T, m_L, m_{\overline L}) = (0.73, 0.29, 0.51) m_{1/2}$ 
and $A_{\yk} = 0.11 m_{1/2}$, with varying 
$m_{1/2}$, and other parameters as described in the text. The lines 
correspond to, from top to bottom, 
$m_{\psi^\pm_1} = 100$, 125, 150, 200, 250, and 400 
GeV (corresponding to
$M_F = 165$, 192, 219, 272, 324, and 
478 GeV respectively). The quantity $M_S$ on the horizontal axis is the 
geometric mean of the new scalar masses. 
The value of $m_{h^0}$ before the corrections is 
taken to be 110 GeV.}
\end{minipage}
\end{figure}

As found in the previous section, the quasi-fixed point behavior of the 
running of the scalar trilinear coupling $a_{\yk}$ implies that the mixing 
parameter 
$X_{\yk}$ is probably actually much smaller than in the ``maximal mixing" case.
Also, the soft supersymmetry breaking squared masses $m_T^2$, $m_L^2$, and
$m_{\overline L}^2$ need not be degenerate. A perhaps better-motivated scenario 
is therefore the gaugino mass dominated case
shown in Figure \ref{fig:delmh}, where I take
$(m_T, m_L, m_{\overline L}) = (0.73, 0.29, 0.51) m_{1/2}$  
and $A_{\yk} = 0.11 m_{1/2}$ 
[see eq.~(\ref{eq:OTLEEmTmL}) and Figure \ref{fig:Akfixed}], with 
$M_T = M_L = M_F$ adjusted so that the lightest new charged fermion mass is
$m_{\psi^\pm_1} = 100$, 125, 150, 200, 250, and 400 GeV.
These results 
were computed using the complete expressions in 
eqs.~(\ref{eq:defMferms}), (\ref{eq:m2neutralscalars})-(\ref{eq:defmatG})
and (\ref{eq:DeltaV}), (\ref{eq:Deltam2h}),
not from the simplified expansion in $\yk$.
The correction to $m_{h^0}$ turns out to be not dramatically sensitive to
$b_T$ and $b_L$ (taken to be 0 here), or $\mu$ (set to $m_{1/2}$ here)
or $\tan\!\beta$ (set to 10 here) provided it is not too small. For a given
value of $m_{\psi^\pm_1}$, the upper bound on corrections to $m_{h^0}$ 
is nearly saturated when $M_T = M_L$.

Figure \ref{fig:delmh} shows that the corrections to $m_{h^0}$ are moderate, 
but can easily exceed 5 GeV for an average new scalar mass $M_S$ less 
than 1 
TeV, provided that 
at least one new charged fermion is lighter than about 200 
GeV. However, it should be kept in mind that the actual corrections
can be larger or smaller than indicated in Figure \ref{fig:delmh},
depending on the details of the new particle spectrum. If the fixed point
behavior for $a_{\yk}$ noted above is evaded somehow, then the corrections
to $m_{h^0}$ could be substantially larger.
The contribution to $\Delta m_{h^0}$ also monotonically increases with the 
scalar masses (for fixed fermion masses), and so in principle could be 
much larger, subject to considerations of fine-tuning that intuitively
should get worse with larger supersymmetry breaking. 
Due to the impossibility of defining an objective measure of fine tuning, 
I will not attempt to quantify the merits of this trade-off, but simply 
note that that even a contribution of a few GeV to $m_{h^0}$ is quite
significant in the context of the supersymmetric little hierarchy 
problem. Smaller fermion masses $m_{\psi_i^0}, m_{\psi^\pm_i}$
may be considered preferred in the sense that this maximizes $\Delta m_{h^0}$.

\section{Precision electroweak effects}
\label{sec:PEW}
\setcounter{equation}{0}
\setcounter{footnote}{1}

The Yukawa couplings $\yk$ and $\yh$ break the custodial symmetry of the 
Higgs sector, and therefore contribute to virtual corrections to $W^\pm$, 
$Z$, and photon self-energies, of the type that are constrained by precision 
electroweak observables. Similarly to the cases analyzed in 
\cite{Martin:2009bg}, these corrections are actually benign, at least if 
one uses $M_t$, $M_W$, and $Z$-peak observables as in the LEP 
Electroweak Working Group analyses \cite{LEPEWWG,LEPEWWG2}. (A different
set of observables are used in \cite{RPP}, leading to
a worse fit.) This is because 
the corrections decouple with larger vector-like masses $M_T$ and $M_L$, 
even if the Yukawa couplings are large and soft supersymmetry breaking 
effects including $m^2_T$, $m^2_L$ and $m^2_{\overline L}$ produce a 
large scalar-fermion hierarchy. In particular, they decouple even when 
the corrections to $m^2_{h^0}$ do not.

The most important new physics contributions to the precision 
electroweak 
observables can be summarized in terms of the 
Peskin-Takeuchi $S$ and $T$ parameters
\cite{Peskin:1991sw} (similar parameterizations of oblique
electroweak observables were discussed in
\cite{Golden:1990ig}).
In this paper, I will use the updated experimental values
\beq
s^2_{\rm eff} &=& 0.23153 \pm 0.00016 \quad \mbox{ref.~\cite{LEPEWWG}}
\\
M_W &=& 80.399 \pm 0.025\>{\rm GeV} \quad \mbox{ref.~\cite{Wmass,LEPEWWG2}}
\\
\Gamma_\ell &=& 83.985 \pm 0.086\>{\rm MeV} \quad \mbox{ref.~\cite{LEPEWWG}}
\\
\Delta \alpha_h^{(5)}(M_Z) &=& 0.02758 \pm 0.00035
\quad \mbox{ref.~\cite{LEPEWWG}}  
\\
M_t &=& 173.1 \pm 1.3\>{\rm GeV} \quad \mbox{ref.~\cite{Tevatrontop}}
\\
\alpha_s(M_Z) &=& 0.1187 \pm 0.0020 \quad \mbox{ref.~\cite{RPP}}
\eeq
with $M_Z = 91.1875$ GeV held fixed. For the Standard Model predictions
for $s^2_{\rm eff}$, $M_W$, and $\Gamma_\ell$ in terms of the other
parameters, I use refs.~\cite{Awramik:2006uz}, \cite{Awramik:2003rn}, and
\cite{Ferroglia:2002rg}, respectively. These values are then used to  
determine the best experimental fit values and the 68\% and 95\% 
confidence level (CL) ellipses for $S$ and $T$, relative to a
Standard Model template with $M_t = 173.1$ GeV and $M_h = 115$ GeV, using  
\beq
\frac{s_{\rm eff}^2}{(s_{\rm eff}^2)_{\rm SM}} &=&
1 + \frac{\alpha}{4 s_W^2 c_{2W}} S - \frac{\alpha c_W^2}{c_{2W}} T
,
\\
\frac{M_W^2}{(M_W^2)_{\rm SM}} &=&
1 - \frac{\alpha}{2 c_{2W}} S + \frac{\alpha c_W^2}{c_{2W}} T
, 
\\
\frac{\Gamma_\ell}{(\Gamma_\ell)_{\rm SM}} &=&
1 - \alpha d_W S + \alpha (1 + s_{2W}^2 d_W) T
,
\eeq
where $s_W = \sin\theta_W$, $c_W = \cos\theta_W$, $s_{2W} =
\sin(2 \theta_W)$, $c_{2W} = \cos(2 \theta_W)$, and
$d_W = (1 - 4 s_W^2)/[(1 - 4 s_W^2 + 8 s_W^4) c_{2 W}]$.
The best fit is found to be $\Delta S = 0.057$ and $\Delta T = 0.080$,
relative to the Standard Model template with 
$M_t = 173.1$ GeV and $M_h = 
115$ GeV. The new physics contributions are given at one-loop order in 
terms of electroweak vector boson self-energy functions $\Pi_{WW}$, 
$\Pi_{ZZ}$, $\Pi_{\gamma\gamma}$, and $\Pi_{Z\gamma}$, which are computed
for the fermions of the $T$, $L$, $\overline{L}$ sector in \STappendix.
The contributions to $S$ and $T$ from scalars will be much 
smaller when
they are much heavier than the fermions, due to decoupling, because 
most of the scalar masses
comes from vector-like soft supersymmetry breaking terms. 
I will therefore neglect those contributions here.
I have also neglected the contributions from the ordinary MSSM superpartners,
which are typically not very large and which also decouple quadratically
with large soft supersymmetry breaking terms.
Note also that
in the OTLEE, TLUDD, and TLEDDD models, the fields that do not have 
Yukawa couplings do not contribute to the $S$ and $T$ parameters.

It is useful to first consider the simple case that electroweak symmetry 
breaking is
treated as a perturbation in the vector-like $T$, $L$, $\overline{L}$ 
sector. For $\yk v_u, \yh v_d \ll M_T, M_L$, I find:
\beq
\Delta T &=& \frac{3 (\yk^2 v_u^2 - \yh^2 v_d^2)^2}{
             32 \pi s_W^2 m_W^2 M_L M_T} f_T(r)
\\
\Delta S &=& \frac{2(\yk^2 v_u^2 + \yh^2 v_d^2)}{5 \pi M_L M_T} f_{S1}(r)
             - \frac{19 \yk \yh v_u v_d}{30 \pi M_T M_L} f_{S2}(r)
\eeq
where $r = M_L^2/M_T^2$, and
\beq
f_T(r) &=& \frac{2\sqrt{r}}{9(1-r)^4} \left [
(6 + 36 r - 36 r^2) \ln(r) + 29 - 36 r - 9 r^2 + 16 r^3 
\right ]
,
\\
f_{S1}(r) &=& \frac{5\sqrt{r}}{6 (r-1)^5} \left [
(1 - 2 r + 9 r^2  - 4 r^3  + 2 r^4 ) \ln(r)
+ 5 r - 6 r^2 + 3 r^3 - 2 r^4 \right ]
,
\\
f_{S2}(r) &=& \frac{10}{19 (r-1)^5} \left [
(4 r + 6 r^2 - 6 r^3 + 8 r^4) \ln(r)
-2 + 21 r - 39 r^2  + 35 r^3  - 15 r^4 \right ]
,
\eeq
normalized so that $f_T(1) = f_{S1}(1) = f_{S2}(1) = 1$. Despite the 
appearances of denominators singular at $r=1$, these functions 
are actually
quite slowly varying.
For the case $\yh = 0$, there follow numerical estimates (for
$r\approx 1$, which very nearly saturates the maximum corrections for
a given value of $M_L M_T$):
\beq
\Delta T &=& 0.42 \left (\frac{\yk}{0.69}\right )^4 \sin^4\!\beta\, 
             \frac{(\mbox{100 GeV})^2}{M_L M_T} 
,
\\
\Delta S &=& 0.18 \left (\frac{\yk}{0.69}\right )^2 \sin^2\!\beta\, 
             \frac{(\mbox{100 GeV})^2}{M_L M_T}
.
\eeq
These contributions decouple quadratically with increasing fermion 
masses, in contrast to the corrections to $m_{h^0}^2$.
However, it should be noted that this expansion in small $\yk v_u$ and 
$\yh v_d$ is
not extremely accurate,
unless both $M_L$ and $M_T$ are much 
larger than 100 GeV, and overestimates the true corrections to $S,T$.

\begin{figure}[!tp]
\begin{minipage}[]{0.49\linewidth}
\caption{\label{fig:ST}
Corrections to electroweak precision observables $S,T$ from the new 
fermions in the $T,L,\overline L$ multiplets, at the fixed point $(\yk, 
\yh) = (0.69, 0)$, for varying $M_T = M_L$. The corrections are evaluated 
using eqs.~(\ref{eq:defS}),(\ref{eq:defT}) and 
(\ref{eq:deltaPIgg})-(\ref{eq:deltaPIWW}). The seven dots on the line 
segment correspond to, from top to bottom, $m_{\psi^+_1} = 100, 125, 150, 
200, 250, 400$ GeV and $\infty$. The experimental best fit is shown as 
the $\times$ at $(\Delta S, \Delta T) = (0.057, 0.080)$. Also shown are 
the $68\%$ and $95\%$ CL ellipses. The 
point $\Delta S= \Delta T = 0$ is defined to be the Standard Model 
prediction for $m_t = 173.1$ GeV and $m_{h^0} = 115$ GeV. Results for 
$M_T \not= M_L$ are very similar; the corrections to $S$ and $T$ are 
slightly smaller than shown here, for the same values of $m_{\psi^+_1}$.}
\end{minipage}
\begin{minipage}[]{0.49\linewidth}
\begin{flushright}
\includegraphics[width=7.7cm,angle=0]{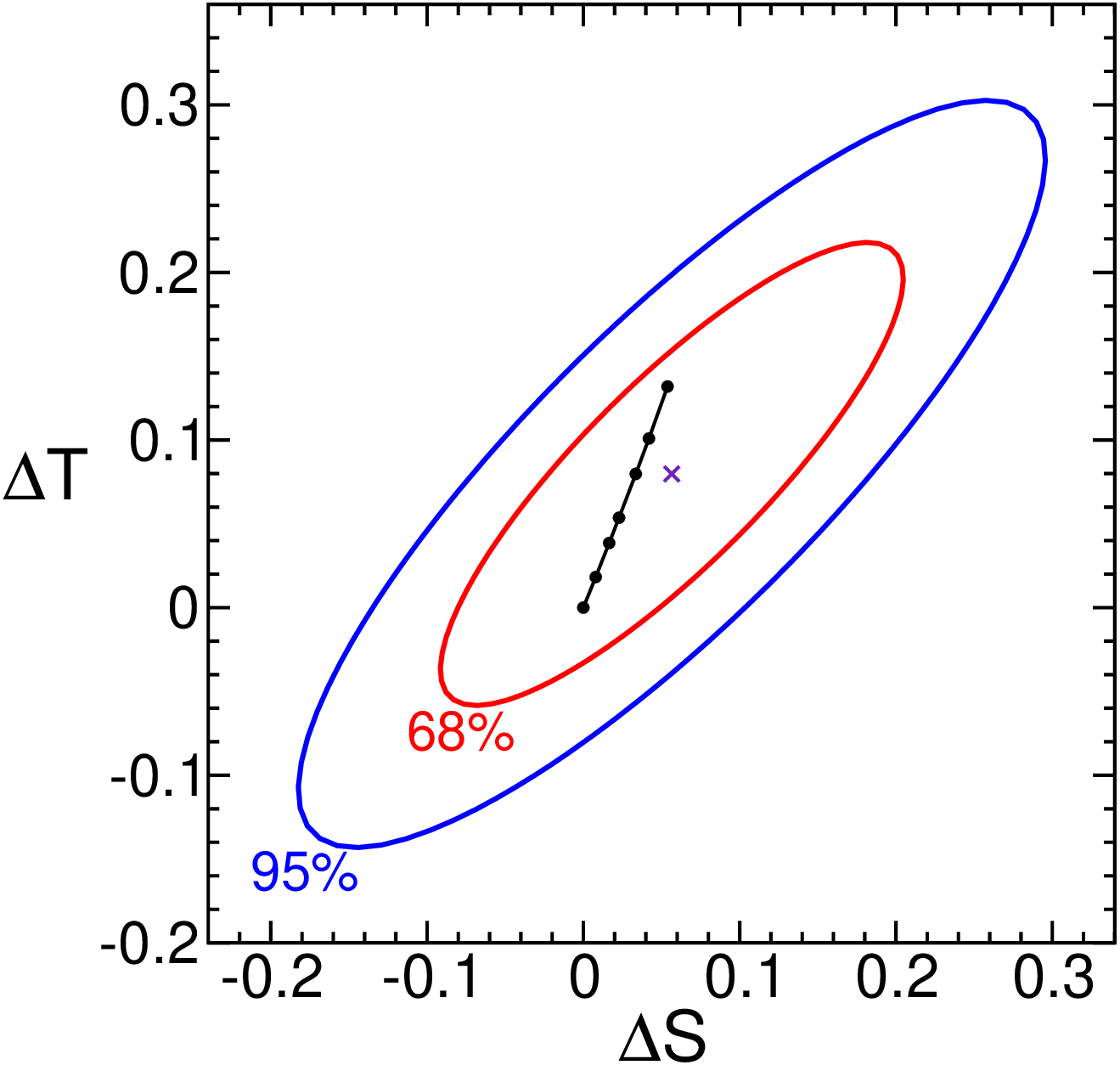}
\end{flushright}
\end{minipage}
\end{figure}

A more accurate evaluation using the complete 
formulas of \STappendix~ is shown in 
Figure \ref{fig:ST}, for the case that $M_T = M_L$ and $\yk = 0.69$,
the quasi-fixed point value.
The seven dots on the line segment correspond to lighter new charged 
fermion masses $m_{\psi^\pm_1} = 100$, 125, 150, 200, 250, 400 GeV and 
$\infty$, from top to bottom. (For reference, the first six points 
correspond to $M_T = M_L = 165$, 192, 219, 272, 324, and 478 GeV 
respectively.) Figure \ref{fig:ST} shows that for charged fermion masses 
not excluded by the CERN LEP2 $e^+e^-$ collider, the $S$ and $T$ 
parameters remain within the current 68\% confidence level ellipse, and 
can even give a better (but not significantly so) fit than the Standard 
Model. Results for $M_T \not= M_L$ are very similar, with corrections to 
$S$ and $T$ that are slightly smaller than shown here, for the same 
values of $m_{\psi^+_1}$. As a caveat, it is important to keep in mind 
that the results above are sensitive to my choice of following the LEP 
Electroweak Working Group \cite{LEPEWWG,LEPEWWG2} in the choice of fit 
observables; if one chose instead the set of observables used in 
\cite{RPP}, the fits to $S$ and $T$ would be worse.

\section{Collider phenomenology of the extra fermions}
\label{sec:BR}
\setcounter{equation}{0}
\setcounter{footnote}{1}

The collider phenomenology of the models discussed in this paper depends 
on the decay modes of the new fermions. First consider the fermions 
$\psi_i^\pm$ and $\psi^0_i$ in the $T$, $L$, and $\overline L$ multiplets 
that are common to the OTLEE, TLUDD, and TLEDDD models. As discussed in 
section \ref{sec:content}, the lightest of these is always the neutral 
fermion $\psi^0_1$. This particle can only decay by virtue of mixing with 
other MSSM fermions. The simplest possibility is that all such mixing is 
forbidden. Then $\psi^0_1$ would be absolutely stable, and in principle 
could be a component of the dark matter. However, the thermal relic 
density would be very low due to an unsuppressed annihilation rate, 
similar to the familiar cases of almost pure wino or higgsino LSPs in the 
MSSM.

If $T$, 
$L$ and $\overline L$ are assigned\footnote{Matter parity
is trivially related to R-parity by an extra factor of $-1$ for 
fermions, and is assumed here to be exactly conserved.
Each of $T$ and $O$, 
and each of the pairs of new superfields 
such as $(L, \overline L)$
that share a superpotential mass term, 
can be independently assigned
either even or odd matter parity, consistently with the requirements of
gauged discrete symmetries \cite{KW,discreteanomaly}.
However, allowing the Yukawa couplings $\yk$ 
and/or $\yh$ requires that $T$,
$L$ and $\overline L$ all have the same matter parity.} 
even matter parity, then the new fermions all have odd R-parity 
and can mix 
with the MSSM charginos and neutralinos through supersymmetric terms 
involving the higgsinos and a supersymmetry-breaking term
involving the winos: 
\beq
W &=& \mu_L L H_u + \mu_{\overline L} \overline L H_d
+ \lambda H_u T H_d,
\\
-{\cal L} &=& M_2' T \widetilde W + {\rm c.c.}
\eeq
These mixing terms enable the decays $\psi_1^0 \rightarrow h^0 \tilde N_1$ 
and
$\psi_1^0 \rightarrow Z^0 \tilde N_1$, and may also enable decays to 
heavier 
ordinary charginos and neutralinos, depending on the 
kinematics. The $M_2'$ term is not one of the 
usual type of soft supersymmetry 
breaking terms, but could arise from a non-renormalizable interaction.
With this even matter parity assignment, the 
supersymmetric mixing terms will 
introduce terms in the scalar potential that will cause 
$L$, $\overline L$ and $T$ to obtain VEVs
(see e.g.~\cite{Espinosa:1991wt}), a possibility not covered in 
the preceding sections and not pursued further here.

\begin{figure}[!tp]
\begin{minipage}[]{0.49\linewidth}
\begin{flushleft}
\includegraphics[width=7.5cm,angle=0]{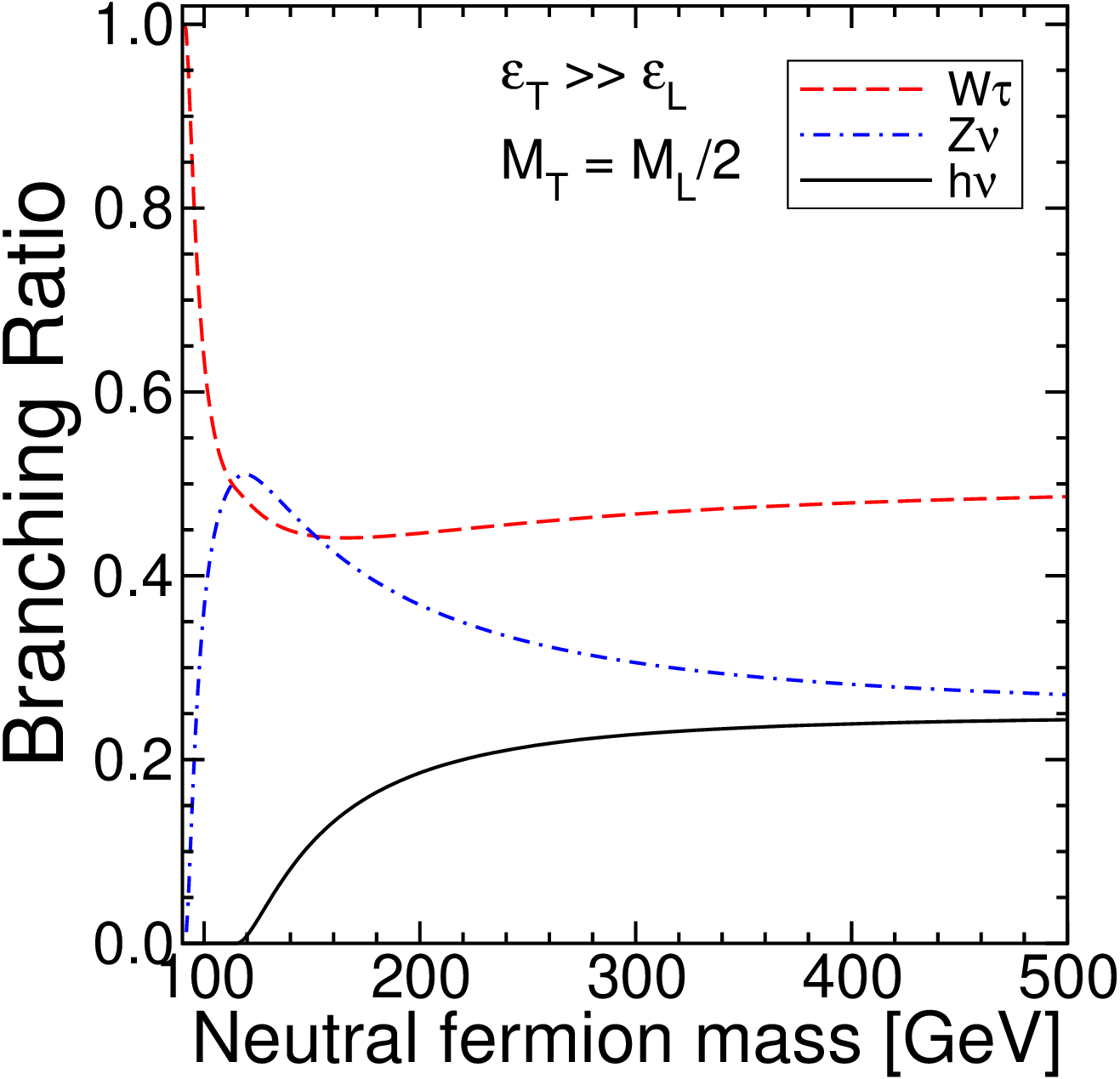}
\end{flushleft}
\end{minipage}
\begin{minipage}[]{0.49\linewidth}
\begin{flushright}
\includegraphics[width=7.5cm,angle=0]{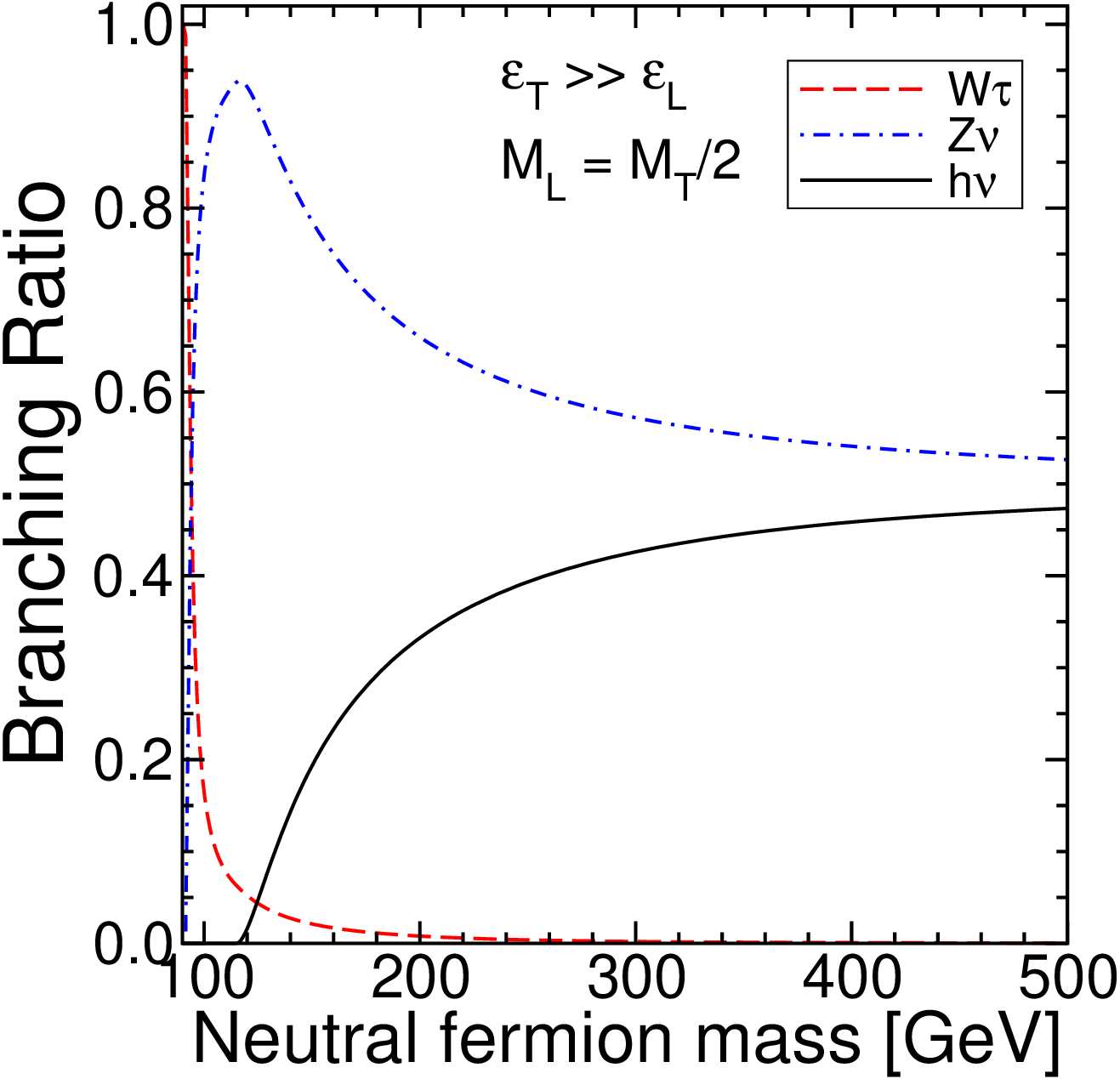}
\end{flushright}
\end{minipage}
\vspace{0.25cm}

\begin{minipage}[]{0.49\linewidth}
\begin{flushleft}
\includegraphics[width=7.5cm,angle=0]{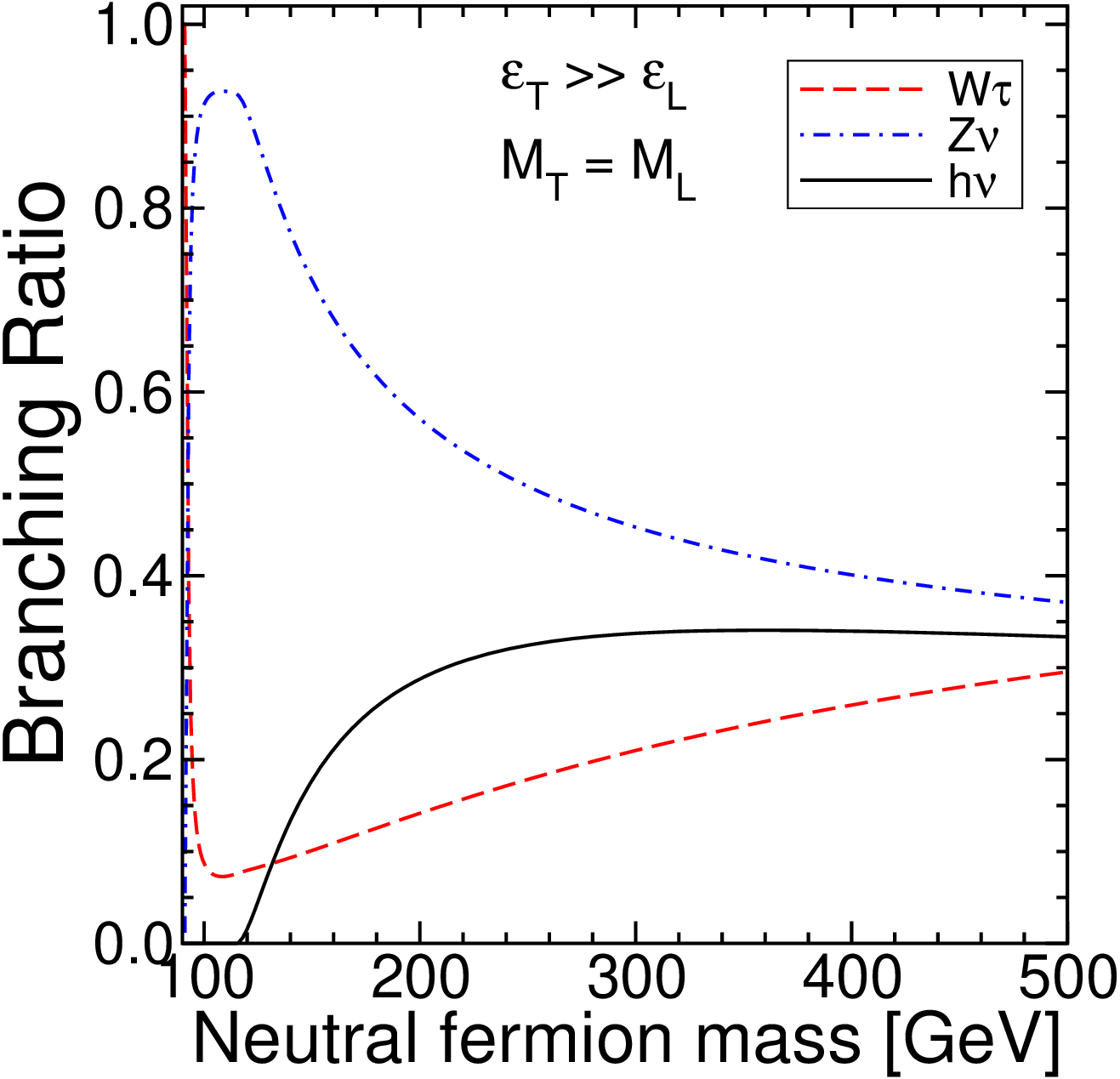}
\end{flushleft}
\end{minipage}
\begin{minipage}[]{0.05\linewidth}
\phantom{xxx.}
\end{minipage}
\begin{minipage}[]{0.44\linewidth}
\caption{\label{fig:decays}
The branching ratios for the lightest (neutral) new fermion $\psi_1^0$,
for decays $\psi_1^0 \rightarrow W\ell$, $Z^0 \nu_\ell$, and $h^0 
\nu_\ell$, 
induced by the mixing in
eqs.~(\ref{eq:defepsTepsL})-(\ref{eq:MpmepsTL}) with $y_\ell \gg \epsT 
\gg \epsL$ and $\tan\!\beta = -\cot\!\alpha = 10$.
Results are shown as a function of the lightest new fermion mass
$m_{\psi_1^0}$
when it is mostly triplet ($M_T = M_L/2$, upper left),
mostly doublet ($M_L = M_T/2$, upper right),
and mixed ($M_T = M_L$, lower left) with $\yk = 0.69$ and $\yh=0$ 
in each case, corresponding to the three cases shown in Figure
\ref{fig:fermionmasses} above. If instead $\epsL \gg \epsT$, then
BR$(\psi_1^0 \rightarrow W\ell) = 1$.
}
\end{minipage}
\end{figure}

If instead $T$,
$L$ and $\overline L$ are assigned odd matter parity, then $\psi^0_1$
has even R-parity
and can only decay through a small mixing with the MSSM leptons, via the 
superpotential terms: 
\beq
W = \epsT H_u T \ell_i + \epsL H_d L \overline{e}_i
.
\label{eq:defepsTepsL}
\eeq
In general this implies lepton family number violation, so it is 
necessary to assume that either the couplings only involve a single
lepton family, or are very small; in addition, lepton universality 
constraints suggest that mixing with the tau lepton may be most important.
In the $(T^0, L^0, \overline L^0, \nu_\ell)$ basis, 
the neutral fermion mass matrix is:
\beq
\widehat{\cal M}_0 = 
\begin{pmatrix}
M_T & \yk v_u & \yh v_d & \epsT v_u\cr
\yk v_u & 0 & -M_L & 0 \cr
\yh v_d & -M_L & 0 & 0 \cr
\epsT v_u & 0 & 0 & 0
\end{pmatrix}
.
\label{eq:M0epsT}
\eeq
The smallest eigenvalue of this will be (a contribution to) 
the squared mass of 
the Standard Model neutrino $\nu_\ell$,
and should be approximately $\epsT^2 v_u^2/M_T $. 
Interpreting the 
bound $\Delta m_{23}^2 < 3 \times 10^{-3}$ eV$^2$ 
\cite{RPP}
as applying to the
$\nu_\ell$ alone (although there could be degeneracy in the neutrino 
sector, so this is not a strict bound), one would arrive at the rough estimate 
$\epsT < 4 \times 10^{-7}\sqrt{M_T/\mbox{100 GeV}}$.
The coupling $\epsL$ is not so constrained, only 
contributing to mixing between the new fermions and the Standard 
Model lepton 
$\ell$ through
the charged fermion mass matrix:
\beq
\widehat{\cal M}_\pm = 
\begin{pmatrix}
M_T & -\sqrt{2} \yh v_d & 0 \cr
\sqrt{2} \yk v_u & M_L & \epsL v_d \cr
\sqrt{2} \epsT v_u & 0 & y_\ell v_d
\end{pmatrix}
.
\label{eq:MpmepsTL}
\eeq
Below I will assume $\epsL, \epsT \ll y_\ell$, so that their effects can 
be treated as small perturbations.
The decays of $\psi^0_1$ due to $\epsT$ and $\epsL$
are: $\psi_1^0 \rightarrow Z\nu_\ell$ and $\psi_1^0 \rightarrow 
Z\overline{\nu}_\ell$
(combined below by an abuse of notation as 
$\psi_1^0 \rightarrow Z\nu$), and $\psi_1^0 \rightarrow h^0\nu_\ell$ 
and $\psi_1^0 \rightarrow h^0\overline{\nu}_\ell$
(similarly combined below as $\psi_1^0 \rightarrow h\nu$),
and
$\psi_1^0 \rightarrow W^\pm \ell^\mp$ (hereafter combined
by writing $\psi_1^0 \rightarrow W \ell$).
The corresponding decay widths are computed in \Decayappendix. 
If $\epsL \gg \epsT$, then the decay will be entirely 
charged current, with BR$(\psi_1^0 \rightarrow W \ell) = 1$. 
Otherwise, the three final states $W \ell$, $Z \nu$, and $h \nu$
can have comparable branching ratios, as shown in Figure 
\ref{fig:decays} for the case $\epsT \gg \epsL$. 
For $m_{\psi_1^0} \lsim m_Z$, the $W\ell$ final state essentially always 
dominates, due to kinematics.
It is notable that in the limit that $\psi^0_1$ is mostly doublet,
the branching ratio to $W\ell$ is suppressed for masses above 100 
GeV. The $\psi_1^0$ decay lengths are necessarily
macroscopic if 
$\epsT < 4 \times 10^{-7} \sqrt{M_T/\mbox{100 GeV}}$
(as suggested by the observed neutrino mass splitting)
and $\epsL = 0$, if $m_{\psi_1^0} \lsim 100$ GeV.
However, the minimum decay length allowed by this condition on $\epsilon_T$
rapidly becomes smaller for larger $m_{\psi_1^0}$, 
with typically $c\tau_{\rm min} \sim$ 1 cm
for $m_{\psi_1^0} = 100$ GeV and $c\tau_{\rm min} \sim$ 1 mm
for $m_{\psi_1^0} = 125$ GeV. 

The heavier new fermions ($\psi_i^\pm$ for $i=1,2$ and $\psi_i^0$ for 
$i=2,3,4$) can decay to the lightest one $\psi_1^0$ via emission of real 
or virtual $W^\pm,Z^0,h^0$ bosons. As can be seen in Figure 
\ref{fig:deltafermionmasses}, the available phase space for the decay of 
$\psi^\pm_1$ to $\psi^0_1$ can be quite small. Depending on $\Delta m$, 
the soft decay products could include one or more pions, a soft jet, or 
more rarely a lepton. These may be hard to detect (and certainly to 
trigger on), especially in a hadron collider environment and especially 
when the lightest fermions are mostly triplet. Extensive studies 
\cite{CDG,Thomas:1998wy,Feng:1999fu,Gherghetta:1999sw,GunionMrenna,Baer:2000bs,Datta:2001hv,Barr:2002ex,Ibe:2006de,Asai:2008sk, 
Buckley:2009kv,Giudice:2010wb} have been made of the somewhat similar 
case of collider production of nearly degenerate MSSM winos or higgsinos. 
When the mass difference for light winos or higgsinos is sufficiently 
small, the $\psi_1^\pm$ decay could manifest itself as a ``stub", a 
stiff, highly ionizing track that ends in the vertex detector or in the 
tracker before making it to the calorimeters, or else has a kink to a 
very soft charged track. It could also be seen as a pion with a 
measurable non-zero impact parameter. These features could be seen in an 
off-line analysis, if the event is triggered by other means.

However, in the present context there are two major differences from the 
MSSM wino and higgsino cases. First, because a large Yukawa coupling 
$\yk$ increases the mass splitting $\Delta m$, making it larger than 
$m_{\pi^\pm}$, the decay of $\psi_1^\pm$ to $\psi_1^0$ will almost always 
occur at a distance scale much smaller than the size of the vertex 
detector or innermost tracker. The critical case of smallest $\Delta m$ 
occurs when the lightest fermions are mostly triplets, where one may take 
over the results of ref.~\cite{CDG}, which show that $c\tau$ is of order 
1 cm (1 mm) for $\Delta m = 0.3$ (0.5) GeV. Comparison with Figure 
\ref{fig:deltafermionmasses} above shows that, for example, $c\tau \gsim$ 
1 cm for $\psi_1^\pm \rightarrow \psi_1^0\pi^\pm$ with $m_{\psi_1^0} < 
200$ GeV will only occur if $M_T \lsim M_L/3$ (assuming that $\yk$ is at 
its fixed point).

The second major difference applies to hadron collider searches. Unlike 
the case of nearly degenerate higgsinos or winos, the new fermions 
described in the present paper are unlikely to occur in significant 
numbers in cascade decays of heavier gluinos, squarks or sleptons, 
because they lack the couplings to MSSM fermion-sfermion pairs that are 
implied by supersymmetry for gauginos and higgsinos. This makes the 
discovery of the new fermions in the case of stable $\psi_1^0$ at hadron 
colliders much more of a challenge in the present case than in those 
studies \cite{CDG,Gherghetta:1999sw,GunionMrenna, 
Baer:2000bs,Barr:2002ex,Asai:2008sk,Giudice:2010wb} that make use of 
gluon, squark or slepton production followed by cascade decays to 
degenerate wino-like or higgsino-like states.

The best existing limits on the new fermions $\psi^\pm_i$ and $\psi^0_i$
come from the LEP2 experiment searches for exotic leptons 
and wino-like and higgsino-like charginos and neutralinos. 
The L3 experiment has produced 
95\% 
confidence level limits 
\cite{L3limits} 
on the mass of a neutral vector-like 
weak doublet fermion ${\psi^0_1}$, 
assuming it decays by $\psi^0_1 \rightarrow \ell W$ with a mean decay length
of less than 1 cm:
\beq
&&m_{\psi^0_1} > \mbox{99.3, 102.7, or 102.6 GeV} \qquad 
(\mbox{for}\,\,\ell = \tau, \mu, \>{\rm or} \> e).
\eeq
L3 also found a limit for the case of a
new charged vector-like doublet 
fermion which decays to a stable neutral partner $\psi_1^0$ according to
$\psi_1^\pm \rightarrow W^{*}\psi^0_1 $
\beq
m_{\psi_1^\pm} > \mbox{102.1 GeV} \qquad (\mbox{for 5 GeV} < 
\Delta m
< \mbox{60 GeV}),
\label{eq:conpsi}
\eeq
where $m_{\psi_1^0} > 40$ GeV is also assumed. As can be seen from Figure 
\ref{fig:deltafermionmasses} above, the assumption on $\Delta m$ in
eq.~(\ref{eq:conpsi}) is indeed 
satisfied when the lighter fermions are mostly doublet and $\yk$ is
at its fixed point value.

However, the preceding limits do not apply to the case where 
the lighter fermions are mostly triplet. 
In particular, the production cross-section 
of $e^-e^+ \rightarrow \psi_1^0 \psi_1^0$ (mediated by the $Z$ boson)
vanishes in the extreme limit of a pure triplet. In that limit, 
$\Delta m$ also becomes very small (see Figure \ref{fig:deltafermionmasses}
above), 
and a different
strategy is needed.
ALEPH and OPAL have searched \cite{ALEPHlimits,OPALlimits} for nearly 
degenerate charged and neutral higgsinos and winos in the 
heavy-sneutrino limit, 
corresponding to the cases $M_T \ll M_L$ and $M_L \ll M_T$ considered 
here (see Figure \ref{fig:fermionmasses} and \ref{fig:deltafermionmasses} 
above).
They used $e^-e^+ \rightarrow \psi_1^+ \psi_1^- \gamma$ and  triggered on 
a hard isolated photon, as suggested in 
refs.~\cite{CDG,Thomas:1998wy,Gherghetta:1999sw}.
The relevant ALEPH limits \cite{ALEPHlimits} for vector-like triplet 
and doublet fermions are:  
\beq
&&m_{\psi_1^\pm} > \mbox{94 GeV} \qquad (\mbox{for triplet with 0.2 GeV} < 
\Delta m < \mbox{5 GeV}),
\label{eq:ALEPHtriplet}
\\
&&m_{\psi_1^\pm} > \mbox{90 GeV} \qquad (\mbox{for doublet with 0.2 GeV} < 
\Delta m < \mbox{5 GeV})
\label{eq:ALEPHdoublet}
\eeq
where $\psi_1^0$ is assumed stable on time scales relevant for collider 
detectors. Comparison with Figure \ref{fig:deltafermionmasses} above shows that
only the triplet limit (\ref{eq:ALEPHtriplet}) is directly relevant for 
our case with $\yk$ large. 
ALEPH also gave different limits for $\Delta m < m_{\pi}$ 
(long-lived charged particle), $\Delta m \approx m_{\pi}$,
and $m_\pi < \Delta m < 0.2$ GeV, 
but the presence of a quasi-fixed point Yukawa coupling 
$\yk$ increases $\Delta m$ well above these ranges. 
The limits obtained by OPAL \cite{OPALlimits}
are similar but slightly weaker.

\begin{figure}[!tp]
\begin{minipage}[]{0.49\linewidth}
\begin{flushleft}
\includegraphics[width=8.0cm,angle=0]{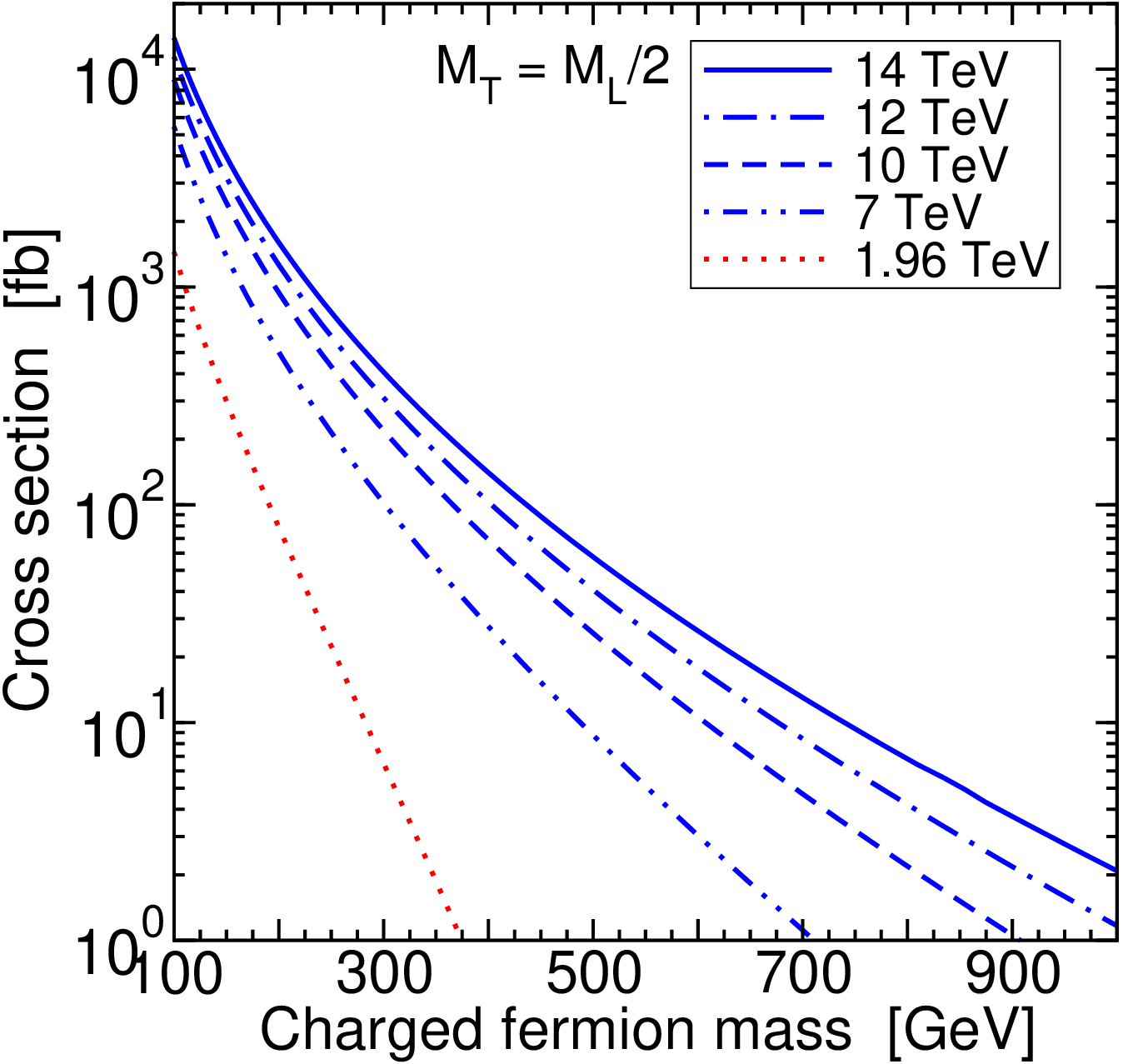}
\end{flushleft}
\end{minipage}
\begin{minipage}[]{0.49\linewidth}
\begin{flushright}
\includegraphics[width=8.0cm,angle=0]{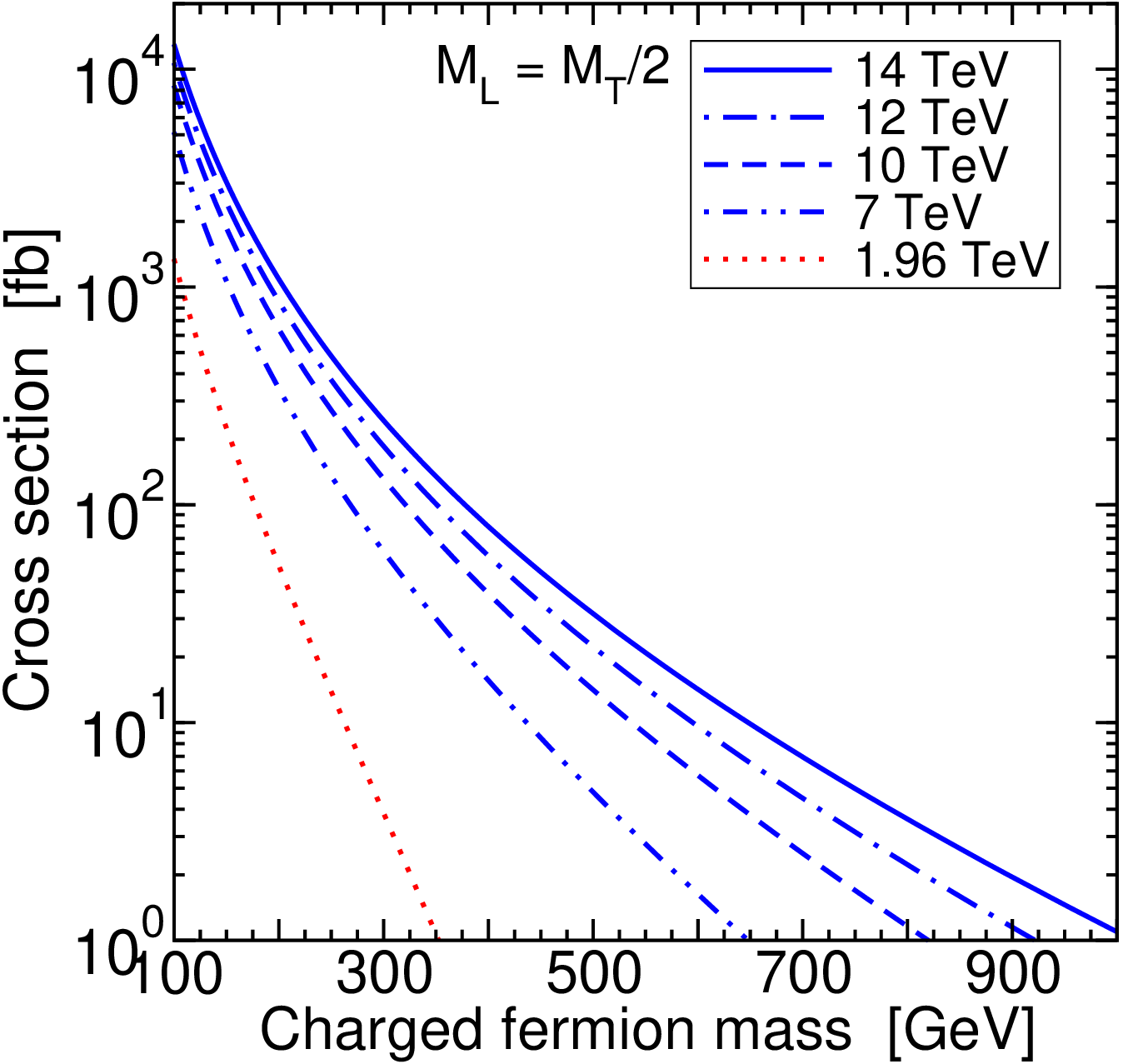}
\end{flushright}
\end{minipage}
\vspace{0.25cm}

\begin{minipage}[]{0.49\linewidth}
\begin{flushleft}
\includegraphics[width=8.0cm,angle=0]{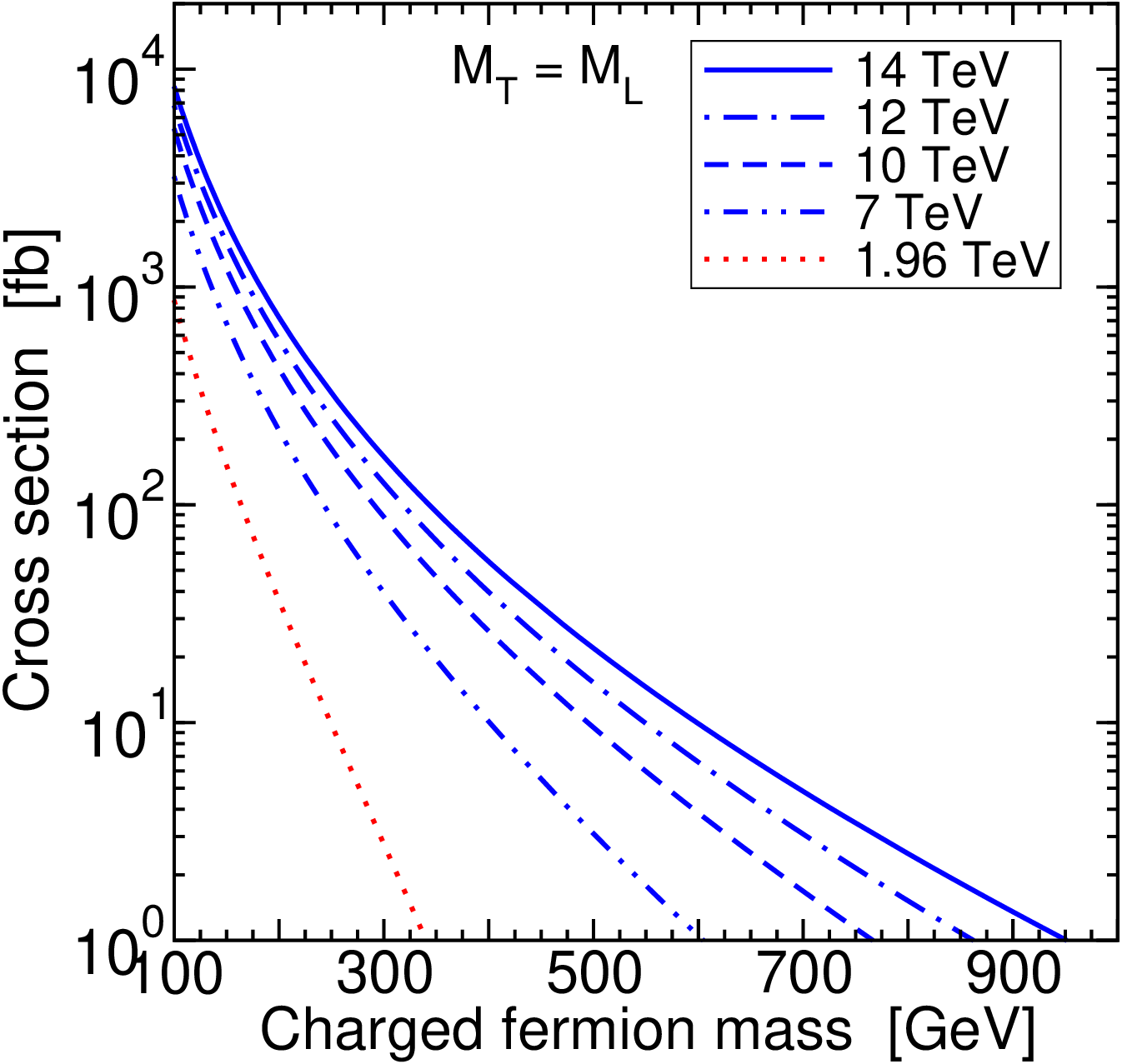}
\end{flushleft}
\end{minipage}
\begin{minipage}[]{0.05\linewidth}
\phantom{xxx.}
\end{minipage}
\begin{minipage}[]{0.44\linewidth}
\caption{\label{fig:crosssections}
The total cross-section for pair production of extra fermions in 
$p\overline p$ collisions at the Tevatron with $\sqrt{s} = 1.96$ TeV (red 
dotted lines), and in $pp$ collisions at the LHC with $\sqrt{s} = 7$, 10, 
12, and 14 TeV. The cross-sections are shown as a function of the lighter 
new charged fermion mass when it is mostly triplet ($M_T = M_L/2$, upper 
left), mostly doublet ($M_L = M_T/2$, upper right), and mixed ($M_T = 
M_T$, lower left) with $\yk = 0.69$ and $\yh=0$ in each case, 
corresponding to the three cases shown in Figure \ref{fig:fermionmasses} 
above. The cross-sections are obtained at leading order (see 
\Crosssectionappendix) and computed using 
CTEQ5LO parton distribution functions \cite{CTEQ5} with $Q$ equal to the 
sum of the masses of the produced particles. 
} 
\end{minipage} 
\end{figure}

The pair production rates of the new fermions for hadron colliders are 
depicted in Figure \ref{fig:crosssections} as a function of
$m_{\psi^\pm_1}$, for the Tevatron
and for four possible LHC energies. The three panels correspond to the
scenarios depicted in Figure \ref{fig:fermionmasses}.
The 
total 
production cross-sections shown in Figure \ref{fig:crosssections} for the 
new fermions at hadron colliders are dominated by the processes:
\beq
p\bar p\>\,\mbox{or}\>\, pp \rightarrow \psi_1^+ \psi_1^-
\>\,{\rm or}\>\, 
\psi_1^\pm \psi_1^0,
\eeq
with $\psi_1^\pm \psi_2^0$ and $\psi_1^0 \psi_2^0$ production making 
smaller but appreciable contributions, especially
when the lightest states have 
significant doublet content. 
There is a larger 
production cross section for $\psi_1^+ \psi_1^0$ than 
for $\psi_1^- \psi_1^0$ at the LHC, because 
there are more $u$ quarks than $d$ quarks in the proton. There are 
evidently no published Tevatron limits that are directly applicable to 
new weakly interacting fermions $\psi_i^\pm, \psi_i^0$ of the type 
discussed in this paper, despite the fact that the production 
cross-section should yield (before any cuts or efficiencies) over $1000$ 
events with the current 8.5 fb$^{-1}$ of integrated luminosity if 
$m_{\psi^\pm_1} < 150$ GeV.

If the $\psi_1^0$ is stable (or quasi-stable on detector length scales), 
then there is a troublesome issue of triggering on the signal events at 
hadron colliders, since the products of $\psi_1^\pm$ decays to $\psi_1^0$ 
will not carry much energy. One possible strategy 
\cite{Feng:1999fu,Ibe:2006de,Buckley:2009kv,Giudice:2010wb} is to 
sacrifice some cross-section and rely on events with one extra hard central 
jet, for example from the parton-level processes
\beq
q\overline q \rightarrow \psi\psi g,\quad
qg \rightarrow \psi\psi q,
\label{eq:extrajet}
\eeq
and then to search for evidence of the $\psi_1^\pm$ track and/or the
decay $\psi_1^\pm \rightarrow \psi_1^0\pi^\pm$ 
in off-line analysis, once the events are triggered on using 
the hard mono-jet.  However, in the present
context, the most striking stub signature
requires $\Delta m \lsim 0.6$ GeV so that $c\tau$ is large enough
to give some events with $\psi_1^0$ making it through at least part of the 
vertex detector or tracker. This can occur, but only 
in the $M_T \ll M_L$ case of
mostly triplet $\psi_1^\pm, \psi_1^0$. For larger $\Delta m$, things will 
probably be much more difficult. 
The severe difficulties involving triggering and backgrounds are 
discussed in detail for the similar situation of
a chargino and neutralino with $\Delta m =$ a few GeV in the case of the 
Tevatron in ref.~\cite{GunionMrenna}.

Another attack \cite{Datta:2001hv,Giudice:2010wb}
on the problem of nearly degenerate winos that should also
be applicable to the present case is
to rely on vector boson fusion, at parton level:
\beq
q q' \rightarrow q q' \psi\psi,
\label{eq:vectorfusion}
\eeq
and then to trigger on the two hard forward quark jets, requiring 
$\ETmiss$ but no (or very soft) hadronic activity in the central region 
from the escaping $\psi_1^0$ or the $\psi^\pm_1$ decays. One can also 
look for soft muons, which occur in 12\% to 20\% of the $\psi^\pm_1$ 
decays for $\Delta m > 0.6$ GeV \cite{CDG}. This method can be applied to 
the cases of larger $\Delta m$ from a few GeV up to tens of GeV. 
Ref.~\cite{Giudice:2010wb} finds a reach of perhaps up to 200 GeV at the 
LHC for 300 fb$^{-1}$ at $\sqrt{s} = 14$ TeV for the comparable case of 
MSSM wino-like fermions, depending on how well the backgrounds can be 
understood.

On the other hand, if $\psi_1^0$
decays promptly to $W\ell$, $Z\nu$, or $h^0\nu$ as discussed above, 
then one can search at hadron colliders for the following 
triggerable final states: 
\beq
&&W^+W^-\ell^+\ell^-,
\qquad
W^-W^-\ell^+\ell^+,
\qquad
W^+W^+\ell^-\ell^-,
\label{eq:WWtautau}\\
&&W^\pm \ell^\mp Z^0 + \ETmiss,\qquad
W^\pm \ell^\mp h^0 + \ETmiss,\\
&&Z^0 Z^0 + \ETmiss,\qquad
Z^0 h^0 + \ETmiss,\qquad
h^0 h^0 + \ETmiss,
\label{eq:hhETmiss}
\eeq
with additional pions or soft jets (or possibly even stubs and high 
impact parameter pions) from the $\psi^\pm_1$ decays to $\psi_1^0$. Here 
$\ell$ could mean any one of $\tau$, $e$, or $\mu$. As discussed above, 
the decays of $\psi_1^0$ are such that only eq.~(\ref{eq:WWtautau}) is 
relevant if $\epsL \gg \epsT$, while otherwise all of 
(\ref{eq:WWtautau})-(\ref{eq:hhETmiss}) are possibilities. The 
same-sign dilepton events in eq.~(\ref{eq:WWtautau}) will occur half of 
the time, due to the Majorana nature of $\psi_1^0$, providing a
low-background signature.\footnote{For
recent detailed studies of the somewhat similar case of same-sign dileptons
from pair production of heavy fourth-family neutrinos at the Tevatron and LHC,
see refs.~\cite{Rajaraman:2010wk,CuhadarDonszelmann:2008jp}.}
However, the leptons 
$\ell$
may well be all (or mostly) taus. In that case, one can still look for
same-sign dilepton events from the leptonic decays of $W$'s and $\tau$'s. 
The discovery potential of the current LHC run
at $\sqrt{s} = 7$ TeV is limited by the aimed-for luminosity of
1 fb$^{-1}$, but from Figure \ref{fig:crosssections} above
there should be several hundred 
events, before cuts and
efficiencies, for $m_{\psi_1^0}$ up to 200 GeV.
There is also an 
intermediate case in which $\psi_1^0$ decays could happen at a measurable 
distance from the beam line 
inside the tracker or the vertex detector, giving an interesting 
signal of two or more charged tracks emanating from a displaced vertex. A 
study of the difficulties and opportunities for discovering the new 
fermions at the Tevatron (where one might hazard a guess that it should 
be possible to set a limit with existing data, for $\psi_1^0$ decaying 
promptly) and LHC in this scenario would be interesting, but is beyond 
the scope of the present paper.

In the OTLEE, TLUDD, and TLEDDD models there are additional fields
whose presence ensures perturbative gauge coupling unification, which
can be searched for at the LHC. They include $SU(2)_L$-singlet
Dirac quarks and leptons, $t'$, $b'$, and $\tau'$, which can decay
only by mixing with their Standard Model counterparts. 
The absence of a GIM-type mechanism
suggests that the mixing with the Standard Model third family is most 
likely to be important, and so one expects decays:
\beq
t' &\rightarrow& Wb,\quad Zt,\quad h^0t,\qquad\mbox{(TLUDD model)}
,
\\
b' &\rightarrow& Wt,\quad Zb,\quad h^0b,\qquad\mbox{(TLUDD and TLEDDD models)}
,
\\
\tau' &\rightarrow& W\nu_\tau,\quad Z\tau,\quad h^0\tau ,
\qquad\mbox{(OTLEE and TLEDDD models).}
\eeq
The branching ratios will depend only on the mass of the new fermion,
as discussed
in Appendix B of ref.~\cite{Martin:2009bg}, and are depicted in the
left panel of Figure 12, the left panel of Figure 13, and
in Figure 14 of that reference. In the limit of large masses,
the $W$, $Z$, and $h^0$ final state branching ratios asymptote to
$0.5$, $0.25$, and $0.25$ respectively, in each case. 
The present Tevatron constraints and LHC search signatures
are also discussed in ref.~\cite{Martin:2009bg} (see also
ref.~\cite{Arnold:2010vs}). 
In the OTLEE model, there is also a color octet fermion $O$, which
can only decay if it has odd R-parity, by virtue of mixing with the 
MSSM gluino through a supersymmetry-breaking Dirac 
mass term $-{\cal L} = M_3' \tilde g O + {\rm c.c.}$ (see 
e.g.~\cite{Polchinski:1982an}).
If this mixing is large enough, then one can have prompt decays of the 
$O$ fermion that are similar to those of the gluino, but with different
kinematics:
\beq
O \rightarrow q\tilde q,\quad q\overline q \widetilde N_i,
\quad q\overline q' \widetilde C_i .
\eeq
Distinguishing the production and decay of the $O$ fermion from that of 
the MSSM gluino could be quite a challenge.
If the mixing is small or absent (if $O$ is assigned even R-parity), then 
the color octet $O$ fermion could
be quasi-stable or stable, and form $R$-hadron-like bound states, with 
signals that have
been well-studied for Tevatron and LHC; see 
\cite{Baer:1998pg}.

\section{Outlook}
\label{sec:outlook}

In general, the constraints implied by measurements of precision 
electroweak observables severely limit the types of new physics that can 
be added to the Standard Model. Minimal supersymmetry gives only small 
corrections to these observables, because it only introduces new 
particles with vector-like (electroweak singlet) masses. Given the 
necessary presence of the vector-like pair $H_u, H_d$ with a bare 
electroweak singlet mass $\mu$ in the MSSM, it is natural to consider 
extensions that contain additional vector-like supermultiplets. In this 
paper, I have studied models with the novel feature of a large Yukawa 
coupling between a new weak triplet and doublet. A motivation for this is 
that, like the models studied in 
\cite{Moroi:1991mg,Babu:2004xg,Babu:2008ge,Martin:2009bg,Graham:2009gy}, 
it can raise the lightest Higgs boson mass. If the new triplet-doublet 
Yukawa coupling and the associated (scalar)$^3$ coupling are at their 
infrared quasi-fixed points, then $m_{h^0}$ can be increased by 5 to 10 
GeV, for an average new scalar mass less than 1 TeV and a lightest new 
charged fermion mass of order 100 to 250 GeV. This increase is 
significant because many otherwise attractive supersymmetric models 
predict that, without such 
a correction, $m_{h^0}$ would be below the LEP2 limit by a similar margin. 
If the couplings are not 
governed by the fixed points, or if the new scalars are heavier,  then 
even larger corrections to $m_{h^0}$ 
are possible.

The new fermions $\psi_i^0$ and $\psi_i^\pm$ in the triplet and doublet 
chiral supermultiplets are subject to direct search limits from LEP2, 
which at most limit their masses to be greater than about 100 GeV. 
Tevatron limits evidently do not yet exist, but may be possible with 
existing data. In principle, there are also indirect constraints from 
precision electroweak observables $S,T$, but these turn out to be easily 
satisfied even if the lightest new charged fermion 
is as light as 100 GeV and the new 
Yukawa coupling $\yk$ is as large as its fixed point value. The lightest 
of the new fermions $\psi_1^0$ could be stable, or could decay through 
mixing with MSSM gauginos, or through mixing with Standard Model leptons. 
In the latter case, the new fermions can be considered as extra vector-like
leptons, and one can have decays $\psi_1^0 \rightarrow W\ell$ or 
$Z\nu$ or $h^0\nu$, with branching ratios and decay lengths that depend 
on the nature of the mixing couplings. At hadron colliders, the 
production of $\psi_1^+\psi_1^-$ and $\psi_1^\pm \psi_1^0$ should 
dominate. If $\psi_1^0$ is stable on collider detector length scales, 
then discovery at the LHC may be difficult, because of small visible 
energy in events and moderate cross sections. If the decay $\psi_1^0$ 
occurs promptly, then one may see same-sign dilepton events with low 
hadronic activity in the central region. In any case, observation of the 
new particles and distinguishing them from supersymmetric backgrounds 
should provide an interesting challenge for the Tevatron and LHC.


\section*{Appendix A: Radiative corrections to new fermion masses}
\label{DeltaMappendix} \renewcommand{\theequation}{A.\arabic{equation}}
\setcounter{equation}{0}
\setcounter{footnote}{1}

The one-loop radiative corrections to the new fermion masses, due to 
diagrams involving electroweak gauge bosons and Higgs bosons, are:
\beq
\Delta m_{\psi_i^0} &=& \frac{1}{16\pi^2 m_{\psi_i^0}} \Bigl \lbrace
|g^Z_{\psi_i^0\psi_j^{0\dagger}}|^2 B_{FV}(\psi_j^0, Z) -
{\rm Re}[(g^Z_{\psi_i^0\psi_j^{0\dagger}})^2] m_{\psi_i^0} m_{\psi_j^0}
B_{\overline{F}V}(\psi_j^0, Z) 
\nonumber \\ &&
+(|g^W_{\psi_i^0\psi_k^{+\dagger}}|^2  + |g^W_{\psi_i^0\psi_k^{-\dagger}}|^2 )
B_{FV}(\psi_k^+, W) -
2 {\rm Re}[g^W_{\psi_i^0\psi_k^{+\dagger}}g^W_{\psi_i^0\psi_k^{-\dagger}}] 
m_{\psi_i^0} m_{\psi_k^+} B_{\overline{F}V}(\psi_k^+, W)
\nonumber \\ &&
+ (|Y^{\Phi^+}_{\psi_i^0\psi_k^-}|^2  +|Y^{\Phi^-}_{\psi_i^0\psi_k^+}|^2)   
B_{FS}(\psi_k^+, \Phi^+)
+ 2 {\rm Re} [Y^{\Phi^+}_{\psi_i^0\psi_k^-}Y^{\Phi^-}_{\psi_i^0\psi_k^+}]
m_{\psi_i^0} m_{\psi_k^+} B_{\overline{F}S}(\psi_k^+, \Phi^+)
\phantom{xxx}
\nonumber \\ &&
+ |Y^{\Phi^0}_{\psi_i^0\psi_j^0}|^2 B_{FS}(\psi_j^0, \Phi^0)
+ {\rm Re}[ (Y^{\Phi^0}_{\psi_i^0\psi_j^0})^2] m_{\psi_i^0} m_{\psi_j^0}
B_{\overline{F}S}(\psi_j^0, \Phi^0)
\Bigr \rbrace
,
\\
\Delta m_{\psi_i^+} &=& \frac{1}{32\pi^2 m_{\psi_i^+}} \Bigl \lbrace
e^2 m^2_{\psi_i^+} [10 - 6 \ln( m^2_{\psi_i^+}/Q^2)]
\nonumber \\ &&
+
(|g^W_{\psi_j^0\psi_i^{+\dagger}}|^2  + |g^W_{\psi_j^0\psi_i^{-\dagger}}|^2 )
B_{FV}(\psi_j^0, W) -
2 {\rm Re}[g^W_{\psi_j^0\psi_i^{+\dagger}}g^W_{\psi_j^0\psi_i^{-\dagger}}] 
m_{\psi_i^+} m_{\psi_j^0} B_{\overline{F}V}(\psi_j^0, W)
\nonumber \\ &&
+
(|g^Z_{\psi_i^+\psi_k^{+\dagger}}|^2  + |g^Z_{\psi_i^-\psi_k^{-\dagger}}|^2 )
B_{FV}(\psi_k^+, Z) -
2 {\rm Re}[g^Z_{\psi_i^+\psi_k^{+\dagger}}g^Z_{\psi_i^-\psi_k^{-\dagger}}] 
m_{\psi_i^+} m_{\psi_k^+} B_{\overline{F}V}(\psi_k^+, Z)
\nonumber \\ &&
+
(|Y^{\Phi^0}_{\psi_i^+\psi_k^-}|^2 + |Y^{\Phi^0}_{\psi_k^+\psi_i^-}|^2)
B_{FS}(\psi_k^+, \Phi^0)
+ 2 {\rm Re}[Y^{\Phi^0}_{\psi_i^+\psi_k^-}Y^{\Phi^0}_{\psi_k^+\psi_i^-}]
m_{\psi_i^+} m_{\psi_k^+} B_{\overline{F}S}(\psi_k^+, \Phi^0)
\nonumber \\ &&
+
(|Y^{\Phi^+}_{\psi_j^0\psi_i^-}|^2 + |Y^{\Phi^-}_{\psi_j^0\psi_i^+}|^2)
B_{FS}(\psi_j^0, \Phi^+)
+ 2 {\rm Re}[Y^{\Phi^+}_{\psi_j^0\psi_i^-}Y^{\Phi^-}_{\psi_j^0\psi_i^+}]
m_{\psi_i^+} m_{\psi_j^0} B_{\overline{F}S}(\psi_j^0, \Phi^+)
\Bigr \rbrace
.\phantom{xxxx}
\eeq
Here the virtual particle labels $j=1,2,3$ and $k=1,2$ and 
$\Phi^0 = h^0, H^0, A^0, G^0$ and $\Phi^+ = H^+, G^+$ are implicitly 
summed over where they appear.
The couplings are given in terms of the mixing matrices defined in 
eqs.~(\ref{eq:defineN}) and 
(\ref{eq:defineUV}) by:
\beq
g^Z_{\psi_i^+ \psi_j^{+\dagger}} &=&
\frac{1}{\sqrt{g^2 + g^{\prime 2}}} \left [
\frac{1}{2} (g^2 - g^{\prime 2}) \Nplus_{i2}^* \Nplus_{j2}
+ g^2  \Nplus_{i1}^* \Nplus_{j1}
\right ]
,
\label{eq:gZpp}
\\ 
g^Z_{\psi_i^- \psi_j^{-\dagger}} &=& 
\frac{1}{\sqrt{g^2 + g^{\prime 2}}} \left [
\frac{1}{2} (-g^2 + g^{\prime 2}) \Nminus_{i2}^* \Nminus_{j2}
- g^2  \Nminus_{i1}^* \Nminus_{j1}
\right ]
,
\\
g^Z_{\psi_i^0 \psi_j^{0\dagger}} &=& 
\frac{1}{2} \sqrt{g^2 + g^{\prime 2}} \left [
\Nzero^*_{i2} \Nzero_{j2} - \Nzero^*_{i3} \Nzero_{j3}
\right ]
,
\\
g^W_{\psi_i^0 \psi_j^{+\dagger}} &=& 
g \left ( \frac{1}{\sqrt{2}} \Nzero^*_{i3} \Nplus_{j2}
- \Nzero^*_{i1} \Nplus_{j1} \right )
,
\\
g^W_{\psi_i^0 \psi_j^{-\dagger}} &=& 
g \Bigl ( \frac{1}{\sqrt{2}} \Nzero^*_{i2} \Nminus_{j2}
+ \Nzero^*_{i1} \Nminus_{j1} \Bigr )
,
\label{eq:gW0m}
\\
Y^{\Phi^0}_{\psi_i^0\psi_j^0} &=& 
-\frac{\yk}{\sqrt{2}} w_{\Phi^0} (N_{i1}^* N_{j2}^* + N_{i2}^* N_{j1}^*)
-\frac{\yh}{\sqrt{2}} x_{\Phi^0} (N_{i1}^* N_{j3}^* + N_{i3}^* N_{j1}^*)
,
\\
Y^{\Phi^0}_{\psi_i^+\psi_j^-} &=& -\yk w_{\Phi^0} \Nplus_{i1}^* 
\Nminus_{j2}^* 
+ \yh x_{\Phi^0} \Nplus_{i2}^* \Nminus_{j1}^* 
,
\\
Y^{\Phi^+}_{\psi_i^0 \psi_j^-} &=& \yk w_{\Phi^+}
(-N_{i1}^* \Nminus_{j2}^* + \sqrt{2} N_{i2}^* \Nminus_{j1}^*),
\\
Y^{\Phi^-}_{\psi_i^0 \psi_j^+} &=& -\yh x_{\Phi^+}
(N_{i1}^* \Nplus_{j2}^* + \sqrt{2} N_{i3}^* \Nplus_{j1}^*) ,
\eeq
with $w_{\Phi^0} = (\cos\!\alpha,\, \sin\!\alpha,\, i \cos\!\beta,\, 
i \sin\!\beta)$
and  $x_{\Phi^0} = (-\sin\!\alpha,\, \cos\!\alpha,\, i \sin\!\beta,\, 
-i \cos\!\beta)$
for $\Phi^0 = (h^0, H^0, A^0, G^0)$, 
and $w_{\Phi^+} = (\cos\!\beta,\, \sin\!\beta)$ 
and $x_{\Phi^+} = (\sin\!\beta,\, -\cos\!\beta)$ for $\Phi^+ = (H^+, G^+)$.
The one-loop self-energy integral functions are, in Feynman gauge and 
following the notation of \cite{Martin:2005ch},
\beq
&&B_{FV}(x,y) = 2 B_{FS}(x,y) = (y-x-s) B(x,y) + A(y) - A(x),
\\
&&B_{\overline{F}V}(x,y) = 
-4 B_{\overline{F}S}(x,y) = 4 B(x,y),
\eeq
where there is an implicit argument $s$ set 
equal to the squared mass of the particle whose mass correction is being 
computed, and 
\beq
A(x) &=& x \ln(x/Q^2) - x,
\label{eq:defAx}
\\
B(x,y) &=& -\int_0^1 dt \ln([t x + (1-t) y - t (1-t) s - i \epsilon]/Q^2)
\label{eq:defBxy}
\eeq
with $Q$ the renormalization scale. 
By convention, the name of a particle appearing as an argument 
of one of these functions stands for the squared mass of the particle. In 
Feynman gauge, $m_{G^0} = m_Z$ and $m_{G^+} = m_W$. The result above is 
similar to that for MSSM charginos and neutralinos in Appendix D of 
\cite{PBMZ} and section V.C of \cite{Martin:2005ch}. In the numerical 
results shown in Figures \ref{fig:fermionmasses} and 
\ref{fig:deltafermionmasses}, I took $m_{H^0}, m_{A^0}, m_{H^+}$ to be 
large enough to decouple, with $\alpha = \beta - \pi/2$, and $\yh=0$, and 
$m_{h^0} = 115$ GeV.


\section*{Appendix B: Contributions to precision
electroweak parameters}
\label{STappendix} \renewcommand{\theequation}{B.\arabic{equation}}
\setcounter{equation}{0}
\setcounter{footnote}{1}

This Appendix gives formulas for the contributions of the fermions in 
the new chiral
supermultiplets $T, L, \overline L$ to the Peskin-Takeuchi precision 
electroweak parameters 
\cite{Peskin:1991sw}.
For convenience I will follow the notations 
and conventions of \cite{Martin:2004id}, which were also followed in 
\cite{Martin:2009bg}. 
The oblique parameters $S$ and $T$ are
defined in terms of electroweak vector boson self-energies by
\beq
\frac{\alpha S}{4 s_W^2 c_W^2} &=& \Bigl [
\Pi_{ZZ}(M_Z^2) - \Pi_{ZZ}(0) 
- \frac{c_{2W}}{c_W s_W} \Pi_{Z\gamma}(M_Z^2) - \Pi_{\gamma\gamma}(M_Z^2)
\Bigr ]/M_Z^2
,
\label{eq:defS}
\\
\alpha T &=& \Pi_{WW}(0)/M_W^2 - \Pi_{ZZ}(0)/M_Z^2
.
\label{eq:defT}
\eeq 
The new
fermion contributions to the electroweak vector boson
self-energies are:
\beq
\Delta \Pi_{\gamma\gamma}(s) &=& 
-\frac{1}{16 \pi^2} 2 g^2 s_W^2 \sum_{i=1,2} G(\psi_i^+),
\label{eq:deltaPIgg}
\\
\Delta \Pi_{Z\gamma}(s) &=& 
-\frac{1}{16 \pi^2} g s_W 
\sum_{i=1,2} (g^Z_{\psi_i^+ \psi_i^{+\dagger}} - 
g^Z_{\psi_i^- \psi_i^{-\dagger}})
G(\psi_i^+)
 ,
\\
\Delta \Pi_{ZZ}(s) &=& 
-\frac{1}{16 \pi^2} \biggl [
\sum_{i,j=1}^3 \left \lbrace |g^{Z}_{\psi^0_i \psi^{0\dagger}_j}|^2 
H(\psi^0_i, \psi^0_j) 
-2 {\rm Re}[(g^{Z}_{\psi^0_i \psi^{0\dagger}_j})^2 ] 
m_{\psi^0_i} m_{\psi^0_j} B(\psi^0_i, \psi^0_j)
\right \rbrace
\nonumber \\ &&
+ \sum_{i,j=1}^2 \Bigl \lbrace (|g^{Z}_{\psi^+_i \psi^{+\dagger}_j}|^2 
+ |g^{Z}_{\psi^-_i \psi^{-\dagger}_j}|^2 ) H(\psi^+_i, \psi^+_j) 
\nonumber \\ &&
-4 {\rm Re}[g^{Z}_{\psi^+_i \psi^{+\dagger}_j}  
g^{Z}_{\psi^-_i \psi^{-\dagger}_j}] 
m_{\psi^+_i} m_{\psi^+_j} B(\psi^+_i, \psi^+_j)
\Bigr \rbrace
\biggr ]
,
\\
\Delta \Pi_{WW}(s) &=& 
-\frac{1}{16 \pi^2} \sum_{i=1}^3 \sum_{j=1}^2
\Bigl \lbrace (|g^{W}_{\psi^0_i \psi^{+\dagger}_j}|^2 
+ |g^W_{\psi^0_i \psi^{-\dagger}_j}|^2 ) H(\psi^0_i, \psi^+_j) 
\nonumber \\ &&
-4 {\rm Re}[g^W_{\psi^0_i \psi^{+\dagger}_j}  
g^W_{\psi^0_i \psi^{-\dagger}_j}] 
m_{\psi^0_i} m_{\psi^+_j} B(\psi^0_i, \psi^+_j)
\Bigr \rbrace
,\phantom{xxx}
\label{eq:deltaPIWW}
\eeq
In these expressions, the couplings are found in 
eqs.~(\ref{eq:gZpp})-(\ref{eq:gW0m}) of 
\DeltaMappendix, and
the one-loop integral functions are given by eqs.~(\ref{eq:defAx}), 
(\ref{eq:defBxy}) and
\beq
H(x,y) &=& \bigl \lbrace [2s -x-y-(x-y)^2/s] B(x,y) 
+ 2 A(x) + 2 A(y) + 2 x + 2 y - 2s/3
\nonumber \\ &&
+ (y-x)[A(x) - A(y)]/s \bigr \rbrace/3,
\\
G(x) &=& H(x,x) + 2 x B(x,x) ,
\eeq 
as in ref.~\cite{Martin:2004id}. 
Particle names should be understood to stand for the 
squared mass when used as an argument of one of these functions, and 
there are implicit arguments $s$ and $Q$ for $B(x,y)$, $H(x,y)$, and $G(x)$
which are identified with
the invariant squared mass argument 
of the self-energy function in which they 
appear and the RG scale.


\section*{Appendix C: Decay widths of new fermions}
\label{Decayappendix} \renewcommand{\theequation}{C.\arabic{equation}}
\setcounter{equation}{0}
\setcounter{footnote}{1}
\setcounter{subsubsection}{0}

This Appendix gives formulas for the decay widths of the lightest new 
fermion in the $T$, $L$, $\overline L$ multiplets, $\psi_1^0$, to 
Standard Model states. These decays are
mediated by the Yukawa couplings $\epsT, \epsL$ in 
eqs.~(\ref{eq:defepsTepsL})-(\ref{eq:MpmepsTL}), 
which provide small mass mixings that 
can be treated as perturbations compared to the other entries in the mass 
matrices. For simplicity, I assume that only one Standard Model lepton 
family $\ell$ is involved.
Define unitary mixing matrices $\nzero$ ($4\times 4$) and 
$\nminus, \nplus$ ($3 \times 3$) by
\beq
\nzero^* \widehat{\cal M}_0 \nzero^\dagger &=& 
{\rm diag}(m_{\nu_\ell}, m_{\psi^0_1}, m_{\psi^0_2}, m_{\psi^0_3})
,
\label{eq:defineNhat}
\\
\nminus^* \widehat{\cal M}_\pm \nplus^\dagger &=&
{\rm diag}(m_{\ell}, m_{\psi^+_1}, m_{\psi^+_2}).
\label{eq:defineUVhat}
\eeq
Then the relevant couplings of $\psi_1^0$ to Standard Model particles 
are:
\beq
g^W_{\psi_1^0 \ell^\dagger} &=& g \bigl (\nzero^*_{21} \nminus_{11} 
+ \frac{1}{\sqrt{2}} \nzero^*_{22} \nminus_{12}
+ \frac{1}{\sqrt{2}} \nzero^*_{24} \nminus_{13} \bigr )
,
\\
g^W_{\psi_1^0 \bar\ell^\dagger} &=& 
g \bigl (-\nzero^*_{21} \nplus_{11} 
+ \frac{1}{\sqrt{2}} \nzero^*_{23} \nplus_{12}
\bigr )
,
\\
g^Z_{\psi_1^0 \nu_\ell^\dagger} &=& \frac{1}{2} \sqrt{g^2 + g^{\prime 2}} 
\bigl (
\nzero_{22}^* \nzero_{12} - \nzero_{23}^* \nzero_{13} + 
\nzero_{24}^* \nzero_{14}
\bigr )
,
\\
Y^{h^0}_{\psi_1^0 \nu_\ell}
&=& 
\frac{\cos\!\alpha}{\sqrt{2}} \bigl [
\yk (\nzero^*_{21} \nzero_{12}^* + \nzero^*_{11} \nzero^*_{22})
+ \epsT (\nzero^*_{21} \nzero_{14}^* + \nzero^*_{11} \nzero^*_{24}) \bigr ]
\nonumber \\ &&
-\frac{\sin\!\alpha}{\sqrt{2}} \yh (\nzero^*_{21} \nzero^*_{13}
+ \nzero^*_{11} \nzero^*_{23} )
.
\eeq
It follows that
\beq
\Gamma(\psi_1^0 \rightarrow W^+ \ell^-) \,=\, 
\Gamma(\psi_1^0 \rightarrow W^- \ell^+) &=&
\frac{m_{\psi_1^0}}{32 \pi} 
(|g^W_{\psi_1^0 \ell^\dagger}|^2 + |g^W_{\psi_1^0 \bar\ell^\dagger}|^2)
(1 - r_W)^2 (2 + 1/r_W)
,
\\
\Gamma(\psi_1^0 \rightarrow Z^0 \nu_\ell) \,=\, 
\Gamma(\psi_1^0 \rightarrow Z^0 \bar \nu_\ell) &=&
\frac{m_{\psi_1^0}}{32 \pi} |g^Z_{\psi_1^0 \nu_\ell^\dagger}|^2 (1 - 
r_Z)^2 (2 + 1/r_Z)
,
\\
\Gamma(\psi_1^0 \rightarrow h^0 \nu_\ell) \,=\, 
\Gamma(\psi_1^0 \rightarrow h^0 \bar \nu_\ell) &=&
\frac{m_{\psi_1^0}}{32 \pi} |Y^{h^0}_{\psi_1^0 \nu_\ell}|^2 (1 - 
r_{h^0})^2 
.\phantom{xx}
\eeq
where $r_X = m^2_X/m^2_{\psi_1^0}$ for $X = W,Z,h^0$, 
and $m_\ell$ and $m_{\nu_\ell}$ are neglected
for kinematic purposes. In the limit of small $\epsT$, the $Z^0\nu_\ell$
and $h^0\nu_\ell$ partial widths go to 0 because $\nzero_{11}$, 
$\nzero_{12}$, $\nzero_{13}$ and $\nzero_{24}$ become small. The dominant
decay in that case is $W\ell$, through the last term in 
$g^W_{\psi_1^0 \bar\ell^\dagger}$.

\section*{Appendix D: Collider production of new fermions}
\label{Crosssectionappendix} \renewcommand{\theequation}{D.\arabic{equation}}
\setcounter{equation}{0}
\setcounter{footnote}{1}

This Appendix contains formulas for the parton-level differential cross 
sections for\footnote{Other parton-level processes with smaller 
cross sections, such as those in eqs.~(\ref{eq:extrajet}) and 
(\ref{eq:vectorfusion}), may turn out to be more relevant for observable 
signals, especially if $\psi_1^0$ is stable. Gluon fusion contributions to
the production, which could be significant for a new chiral family of leptons
\cite{Willenbrock:1985tj}, should be very small in the vector-like case.} $u\overline d \rightarrow 
\psi_i^+ \psi_j^0$ and $f \overline f \rightarrow \psi_i^- \psi_j^+$ and 
$f \overline f \rightarrow \psi_i^0 \psi_j^0$. The general form of the 
result is
\beq
\frac{d\sigma}{dt} = \frac{1}{64 N_c \pi s^2} \sum |{\cal M}|^2
\eeq
with $\sum |{\cal M}|^2$ to be given below for each process, and
\beq
t = [m_i^2 + m_j^2 - s + \lambda^{1/2}(s, m_i^2, m_j^2) \cos\!\theta]/2
\eeq
where $\theta$ is the angle between the first initial-state fermion and 
the outgoing fermion labeled $i$, and $\sqrt{s}$ is the 
center-of-momentum energy, and $\lambda(x,y,z) = x^2 + y^2 + z^2 - 2 x y 
- 2 x z - 2 y z$. Also $s+t+u = m_i^2 + m_j^2$. Couplings needed below 
are found in \DeltaMappendix. For quarks (leptons) in the initial state, 
$N_c = 3$ (1). The computations and presentations here are similar to
those in 
sections 6.12, 6.14, and 6.15 of \cite{DHM}.

For $u \bar d \rightarrow \psi_i^+ \psi_j^0$, the result is
\beq
\sum |{\cal M}|^2 = |c_1|^2 (u - m_{\psi_i^+}^2)(u - m_{\psi_j^0}^2)
+ |c_2|^2 (t - m_{\psi_i^+}^2)(t - m_{\psi_j^0}^2)
+ 2 {\rm Re}[c_1 c_2^*] s m_{\psi_i^+} m_{\psi_j^0} 
,
\eeq
where
\beq
c_1 = \frac{\sqrt{2} g}{s - m_W^2} g^W_{\psi_j^0 \psi_i^{+\dagger}}, 
\qquad\quad
c_2 = -\frac{\sqrt{2} g}{s - m_W^2} \bigl (g^W_{\psi_j^0 
\psi_i^{-\dagger}} \bigr )^*.
\eeq

For $f \bar f \rightarrow \psi_i^- \psi_j^+$, the result is
\beq
\sum |{\cal M}|^2 &=& 
(|c_1|^2 + |c_4|^2) (u - m_{\psi_i^-}^2)(u - m_{\psi_j^+}^2)
+
(|c_2|^2 + |c_3|^2) (t - m_{\psi_i^-}^2)(t - m_{\psi_j^+}^2)
\nonumber \\ &&
+
2 {\rm Re}[c_1 c_2^* + c_3 c_4^*] s m_{\psi_i^-} m_{\psi_j^+} ,
\eeq
where
\beq
c_1 &=& -\frac{2Q_f e^2}{s} \delta_{ij} +
\frac{2(T^3_f - Q_f s_W^2)g}{c_W 
(s-m_Z^2)}g^Z_{\psi_j^-\psi_i^{-\dagger}},
\\
c_2 &=& -\frac{2Q_f e^2}{s} \delta_{ij} -
\frac{2(T^3_f - Q_f s_W^2)g}{c_W (s-m_Z^2)} 
g^Z_{\psi_i^+\psi_j^{+\dagger}},
\\
c_3 &=& -\frac{2Q_f e^2}{s} \delta_{ij} -
\frac{2 Q_f s_W^2 g}{c_W (s-m_Z^2)} g^Z_{\psi_j^-\psi_i^{-\dagger}},
\\
c_4 &=& -\frac{2Q_f e^2}{s} \delta_{ij} +
\frac{2 Q_f s_W^2 g}{c_W (s-m_Z^2)} g^Z_{\psi_i^+\psi_j^{+\dagger}},
\eeq
with $(Q_f, T^3_f) = (2/3, 1/2)$ for $f = u$, and
$(-1/3, -1/2)$ for $f = d$, and
$(-1, -1/2)$ for $f=e$.

For $f \bar f \rightarrow \psi_i^0 \psi_j^0$, 
\beq
\sum |{\cal M}|^2 &=&
(|c_1|^2 + |c_2|^2) [
 (u - m_{\psi_i^0}^2)(u - m_{\psi_j^0}^2)
 +
 (t - m_{\psi_i^0}^2)(t - m_{\psi_j^0}^2)]
\nonumber \\ &&
-
2 {\rm Re}[c_1^2 + c_2^2] s m_{\psi^0_i} m_{\psi^0_j} ,\phantom{xxxxx}
\eeq
where
\beq
c_1 =
\frac{2 (T^3_f - Q_f s_W^2) g}{c_W (s-m_Z^2)} 
g^Z_{\psi_i^0\psi_j^{0\dagger}}
,\qquad
c_2 =
-\frac{2 Q_f s_W^2 g}{c_W (s-m_Z^2)} 
g^Z_{\psi_i^0\psi_j^{0\dagger}}.
\eeq
When $i=j$, one must also include an additional factor of $1/2$
in the total cross section
for identical final state particles.

\bigskip \noindent 
{\it Acknowledgments:} 
This work was supported in part by the National Science Foundation grant 
number PHY-0757325.


\end{document}